\documentclass[12pt]{article}

\usepackage{multirow}
\usepackage{multicol}
\usepackage{cite}

\usepackage[nointlimits,reqno]{amsmath}
\usepackage{amssymb}
\usepackage{slashed}
\numberwithin{equation}{section} 

\usepackage{graphicx}

\usepackage{adjustbox}
\usepackage{paralist, tabularx}
\usepackage[toc,page]{appendix}
\usepackage{stmaryrd}
\usepackage{textcomp}

\usepackage{subcaption}
\usepackage{float}
\usepackage{caption}
\usepackage[pdftex,
 pdfproducer=TeXShop,
 pdfcreator=pdflatex]{hyperref}

\usepackage{luty}

\newcommand{\met}{\Sla{E}_{T}}
\newcommand{\pt}{\ensuremath{p_{T}}}

\begin{document}

\newcommand{\norm}[1]{|\!\gap|#1|\!\gap|}
\newcommand{\aavg}[1]{\avg{\!\avg{#1}\!}}
\newcommand{\ZZ}{\mathbb{Z}_2}

\begin{titlepage}

\title{Almost Inert Higgs Bosons\\
at the LHC}

\author{Christina Gao,\ \ 
Markus A. Luty}

\address{Center for Quantum Mathematics and Physics (QMAP)\\
University of California, Davis, California 95616}

\author{Nicol\'as A. Neill}

\address{Department of Physics and CCTVal\\
Universidad T\'ecnica Federico Santa Mar\'ia, Valpara\'iso  2340000, Chile}

\begin{abstract}
Non-minimal Higgs sectors are strongly constrained by the agreement of
the measured couplings of the 125~GeV Higgs with Standard Model predictions.
This agreement can be explained by an approximate $\ZZ$ symmetry under which 
the additional Higgs bosons are odd.
This allows the additional Higgs bosons to be approximately inert, meaning 
that they have suppressed VEVs and suppressed mixing with the Standard Model Higgs.
In this case, single production of the new Higgs bosons is suppressed,
but electroweak pair production is unsuppressed.
We study the phenomenology of a minimal 2 Higgs doublet model that realizes
this scenario. 
In a wide range of parameters, the phenomenology of the model is
essentially fixed by the masses
of the exotic Higgs bosons, and can therefore be explored systematically.
We study a number of different plausible signals in this model, and show
that several LHC  searches can constrain or discover additional Higgs bosons
in this parameter space.
We find that the reach is significantly extended at the high luminosity LHC.
\end{abstract}
\end{titlepage}

\section{Introduction}
The discovery of the 125~GeV Higgs boson at the LHC \cite{Aad:2012tfa,Chatrchyan:2012xdj}
has been rapidly followed by an impressive program of measurement of Higgs couplings
that tells us that the Higgs couplings are consistent with Standard Model
predictions at the 10$\%$ level 
\cite{Khachatryan:2016vau,ATLAS:2018doi,Sirunyan:2018koj}.
Further improving the Higgs coupling measurements
is an important part of the ongoing physics program at the LHC and future colliders.
An important complementary probe of the Higgs sector are direct
searches for additional Higgs bosons.
Additional Higgs multiplets are intrinsic to 
many extensions of the Standard Model that address the
problem of naturalness, such as supersymmetry or composite Higgs models.
In addition, from a purely phenomenological point of view, it is important to 
experimentally constrain non-minimal Higgs sectors that could play a role 
in electroweak symmetry breaking
and the generation of elementary particle masses without reference to 
specific models of naturalness.

The consistency of the observed Higgs couplings with the Standard Model strongly
constrain the possibilities for discovery of additional Higgs bosons.
The simplest explanation for this consistency is that any additional Higgs multiplets
have large positive electroweak-preserving mass terms.
These models have a ``decoupling limit'' where the 
quadratic terms of the new Higgs fields get large, with other couplings held fixed\cite{Haber:1989xc,Gunion:2002zf}.
In this limit, the physical masses of the new Higgs bosons becomes large,
and their effects decouple at low energies.
Probing additional Higgs bosons near the decoupling limit
is therefore very difficult.

Another limit of multi-Higgs models that is often studied in the literature
is the ``alignment limit'' where the lightest $CP$ even physical Higgs boson 
$h$ is closely aligned with the VEV in the multi-Higgs field space \cite{Gunion:2002zf,Craig:2013hca}.
The decoupling limit implies the alignment limit, but alignment does not require 
the new Higgs bosons to be heavy.
Alignment without decoupling is not guaranteed by any symmetry, and is therefore
an accidental (or fine-tuned) property of the Higgs potential.
The alignment limit has a distinctive phenomenology.
The approximate alignment of the 125~GeV mass eigenstate $h$ with the Higgs VEV
guarantees that the couplings $hVV$ ($V = W, Z$) are close to the
Standard Model values.
Since these are among the most precisely measured Higgs couplings, this 
partially explains the Standard-Model-like nature of the observed Higgs bosons.
The alignment limit implies that couplings of the form $HVV$ are suppressed, 
where $H$ denotes a new Higgs boson.
However, the couplings $Hff$ ($f = \text{fermion}$) are allowed to be 
unsuppressed, so one searches for signals involving the heavies fermions
$t$, $b$, and $\tau$ \cite{Delgado:2013zfa,Craig:2013hca,Carena:2013ooa}.

In this paper we consider a simple symmetry explanation for the 
Standard-Model like couplings of the 125~GeV Higgs that allows additional
Higgs bosons to be light.
We assume that there are additional Higgs doublets that are odd under an 
approximate $\ZZ$ symmetry, while all Standard Model fields (including
the Standard Model Higgs doublet) are even under $\ZZ$.
We first consider the limit where the $\ZZ$ symmetry is exact,
and then include small explicit breaking.
First, note that the Yukawa couplings of the additional Higgs doublets
to Standard Model fermions are forbidden by $\ZZ$ symmetry.
Next we consider the couplings of the Higgs bosons to vector bosons.
We assume that the $\ZZ$ odd Higgs fields have positive
quadratic terms, so that they have vanishing VEV,
and the $\ZZ$ symmetry is not spontaneously broken.
In this case the $\ZZ$-even and $\ZZ$-odd Higgs bosons do not
mix, and the Standard Model Higgs doublet is entirely responsible for electroweak 
symmetry breaking.
In this case, the vector couplings of the $\ZZ$ even Higgs boson are the
same as in the Standard Model, so the $\ZZ$ symmetry gives a limit
where the Higgs is naturally Standard Model-like.
In this scenario, the additional Higgs bosons are called ``inert'' because they 
do not contribute to electroweak symmetry breaking \cite{Deshpande:1977rw,Ginzburg:2010wa}.
In the inert limit, the lightest $\ZZ$ odd particle is stable, and may be
dark matter \cite{Ma:2006km,Barbieri:2006dq,LopezHonorez:2006gr,Goudelis:2013uca,Arhrib:2013ela,Ilnicka:2015jba,Ilnicka:2018def,Belyaev:2016lok,Kalinowski:2018ylg}.

We consider the case where the $\ZZ$ symmetry is approximate,
so the new Higgs bosons are only approximately inert. 
We will assume that all $\mathbb{Z}_2$ breaking terms are suppressed by a small
dimensionless parameter $\ep$.
The parameter $\ep$ then suppresses single production of the new Higgs bosons,
as well as their decays.
Therefore, any deviation of the 125~GeV couplings to vectors or fermions from the
Standard Model prediction is suppressed by $\ep$, and the observed Higgs
is naturally Standard Model-like.
\begin{figure}[t]
\centering
\includegraphics[width=0.6\textwidth]{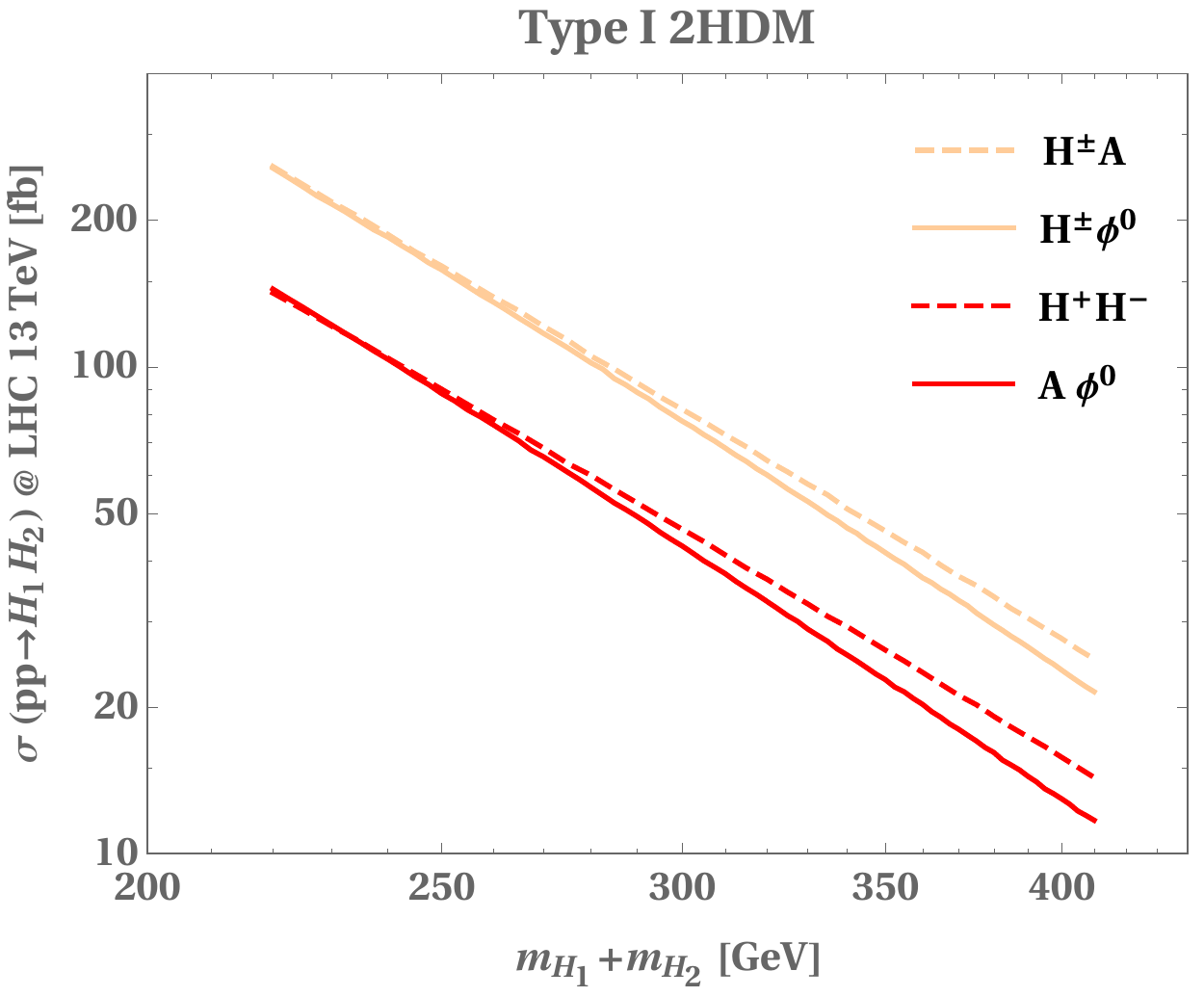}
\caption{Production cross section for pairs of exotic Higgs bosons as a function of the total mass of the final state at LHC 13 TeV.
Each curve corresponds to $\si^{\mathrm{LO}}_{pp\shortrightarrow V \shortrightarrow H_1 H_2}$, where $V=W^\pm,Z$ as appropriate.
For $H^\pm A$ we consider $m_{H^\pm}=m_A$. 
For $H^\pm \phi^0$ and $A\phi^0$ the cross section depends on two independent masses;
we choose $m_{H^\pm}=m_A = 110\mbox{ GeV}$ and vary $m_{\phi^0}$.
The cross sections were obtained with MadGraph \cite{Alwall:2014hca}.
}\label{cross-section}
\end{figure}

The focus of this paper is on the collider signatures of these ``almost inert''
Higgs bosons.
Standard searches for exotic Higgs particles at the LHC rely on single
production of the Higgs particles, which is suppressed by $\ep$ in
this scenario.
For moderate values of $\ep$ (roughly $\ep \lsim 0.1$) these searches
are completely ineffective due to low production cross-sections.
However, couplings of the form $VHH$ ($V = W, Z, \ga$, $H = \text{exotic Higgs}$)
are fixed by gauge invariance, and are unsuppressed in the inert limit.
These are therefore the main production mode for the new Higgs particles.
The decay of the new Higgs bosons to Standard Model particles is also 
suppressed by $\ep$.
This means that heavier $\ZZ$ odd Higgs bosons will preferentially decay
weakly to lighter $\ZZ$ odd Higgs bosons, followed by a slower decay of the
lightest $\ZZ$ odd particle to Standard Model particles.
This leads to cascade decays with multiple Standard Model particles in
the final states.
Although the last stage of the decays is suppressed by $\ep$, it will still
be prompt as long as  $\ep \gsim 10^{-4}$ (for masses of the additional scalars $\gsim$ 200 GeV).\footnote{See Table (\ref{tablectau}) for a numerical example of the values of $\epsilon$ required to have prompt decays for different masses of the additional scalars.}
Thus, for many orders of magnitude in the $\ZZ$ breaking parameter
($10^{-4} \lsim \ep \lsim 10^{-1}$) the phenomenology
is dominated by prompt cascade decays.
For $\epsilon$'s less than $\mathcal{O}(10^{-4})$, the approximately $\ZZ$-odd scalars become long-lived, so displaced vertices searches can be relevant in this regime (e.g. \cite{Aaboud:2018jbr,Khachatryan:2016sfv,CMS:2016ybj}).
Here we limit ourselves to the prompt case, leaving the potential of displaced vertices searches for future work.

In fact, the phenomenology of this model is
almost completely determined by the masses of the $\ZZ$-odd Higgs particles, 
i.e., the charged Higgs $H^\pm$, the neutral CP-even Higgs $\phi^0$ and the neutral CP-odd Higgs $A$.
The $\mathbb{Z}_2$ symmetry allows
electroweak symmetry violating mass splittings within the additional Higgs
multiplets.
(These arise from $\mathbb{Z}_2$ invariant terms in the Higgs potential
such as $|H_1^\dagger H_2|^2$,
where $H_{1,2}$ are the $\ZZ$ even and odd
Higgs doublets, respectively.)
The leading production process for the new Higgs bosons is pair production
from a virtual $W$, $Z$, or $\ga$, namely\footnote{We denote the Standard Model Higgs doublet by $h^0$ and the $CP$-even $\ZZ$ odd Higgs boson by $\phi^0$.}
\[
W^* \to H^\pm \phi^0,
\quad
W^* \to H^\pm A,
\quad
Z^*/\ga^* \to H^+ H^-,
\quad
Z^* \to A \phi^0.
\] 
The production rate for these processes is fixed by gauge invariance,
and the rates at the LHC are shown in Fig.~\ref{cross-section}.
The heavier new Higgs particles will generically 
have cascade decays to lighter members of the new Higgs multiplet by 
emitting a (possibly virtual) $W$ or $Z$.
These decays are not suppressed by $\ep$, and therefore generically
dominate over decays to Standard Model states.
The lightest additional Higgs then has a ``slow'' decay only through
$\mathbb{Z}_2$ violating couplings.
These can be thought of as arising from mixing with the Standard Model
Higgs, and therefore go to the heaviest kinematically accessible
Standard Model state.
This gives rise to a rich set of many-particle final states featuring
the heaviest Standard Model particles: $t$, $h$, $Z$, $W$, $b$, and $\tau$.

The decay cascades are generally dominated by a single decay mode
at each stage of the decay, so the signal is determined
completely by the masses of the new Higgs bosons.
The lightest $\ZZ$ odd Higgs boson decays to the heaviest kinematically
available Standard Model particles.
Weak production of $\ZZ$ odd Higgs bosons can give
$H^{\pm}A^0$,
$H^+ H^-$, $\phi^0 H^{\pm}$, or
$\phi^0 A^0$. 
These then cascade decay down to the lightest $\ZZ$ odd
Higgs boson, generating a state with one or more vector bosons
($W$ and/or $Z$) plus $\phi^0 \phi^0$, $H^+ H^-$ or $A^0 A^0$.
The lightest $\ZZ$-odd Higgs boson then decays to Standard Model
particles.
Because these decays occur {\it via} mixing with the Standard Model
Higgs, these decays are to the heaviest kinematically accessible
Standard Model final state.
\begin{figure*}[t]
\centering
\includegraphics[scale=1]{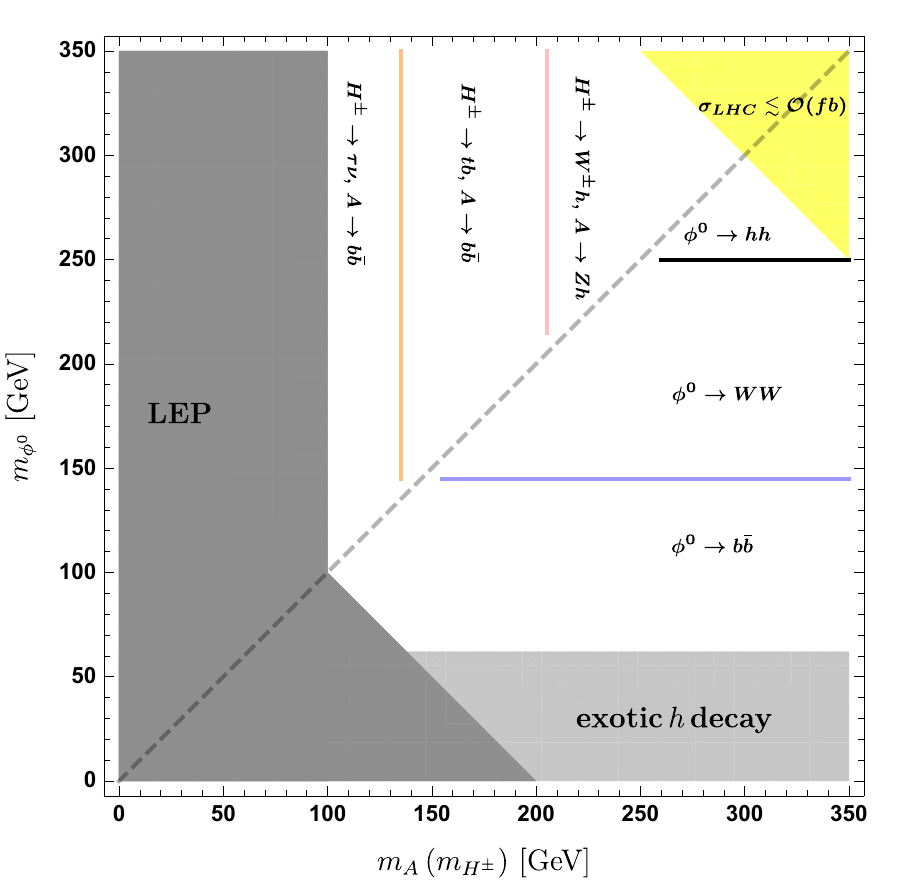}
\caption{Dominant decay modes for $H^{\pm}(A)$ and $\phi^0$, 
assuming in each case it is the lightest $\ZZ$ odd Higgs boson.
A rough estimate of LEP bounds are shown in dark grey.
See text for additional details.}
\label{decay}
\end{figure*}
These decays are summarized in Fig.~\ref{decay},
which also shows an estimate of the region where LEP is sensitive to the $\ZZ$-odd Higgs bosons.
LEP can directly produce $\phi^0A$ and $H^+H^-$ 
{\it via} $Z^*/\gamma^*$, so it can probe the region where these states are kinematically available. The actual limits (see Refs.~\cite{Schael:2006cr,Abbiendi:2013hk}) are slightly weaker than the estimate in the figure.
Other constraints might come from $h\to \gamma\gamma$. Charged Higgs loop can potentially give a large contribution to this decay. As explained in the appendix, the almost inert Higgs corresponds to a large $\tan\beta$ limit of the type-I 2HDM.
As shown in Ref. \cite{Arbey:2017gmh}, in this limit there are no any other constraints excepting the ones from LEP.
In addition, Fig.~\ref{decay} gives a rough indication of the LHC reach for this model by showing the parameter space where the LHC production rate for a pair of
$\ZZ$-odd Higgs bosons becomes smaller than $\sim 1$~fb.
We also restrict ourselves to masses of $\phi^0$ in the range
\[
62.5~\text{GeV} < m_{\phi^0} < 250~\text{GeV}
\]
to avoid the processes $\phi^0\to hh$ and $h\to \phi^0\phi^0$.
Processes involving $\phi^0\to hh$ will be very challenging due
to the low rate.
The process $h \to \phi^0\phi^0$ can become important when it is kinematically accessible, therefore it is constrained by exotic $h_{SM}$ decays \cite{Sirunyan:2018mot,Aaboud:2018esj,Sirunyan:2018owy,Aaboud:2019rtt}.
Moreover, the $h \to \phi^0\phi^0$ decay width depends on the parameters of the full Higgs potential.
Given that in this work we wish to investigate the phenomenology dictated by introducing a small $\ZZ$-breaking effect, we leave this model-dependent channel for future work.

We focus on the white region in Fig.~\ref{decay}, which illustrates the
parameter space we are probing.
The fact that this parameter space can be represented on a
2-dimensional plot means that the phenomenology 
of this scenario can be explored systematically.

\begin{figure*}[t!]
    \centering
     \begin{subfigure}[t]{0.48\textwidth}
    \centering
        \includegraphics[scale=0.9]{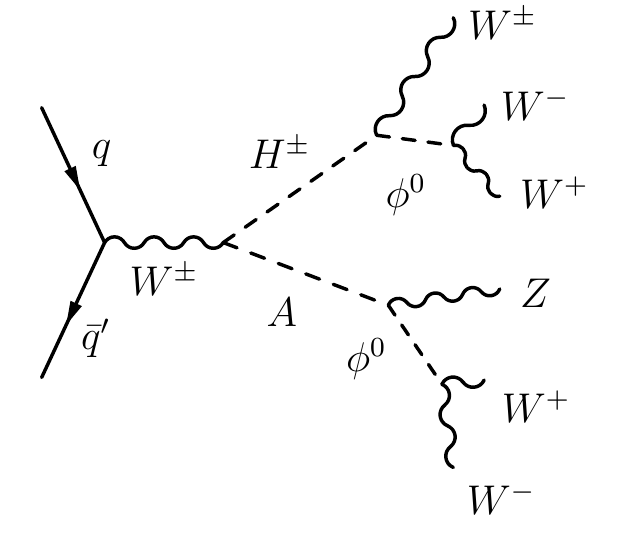}
        \caption{}
        \label{3l_feyn}
    \end{subfigure} 
     \begin{subfigure}[t]{0.48\textwidth}
    \centering
        \includegraphics[scale=0.9]{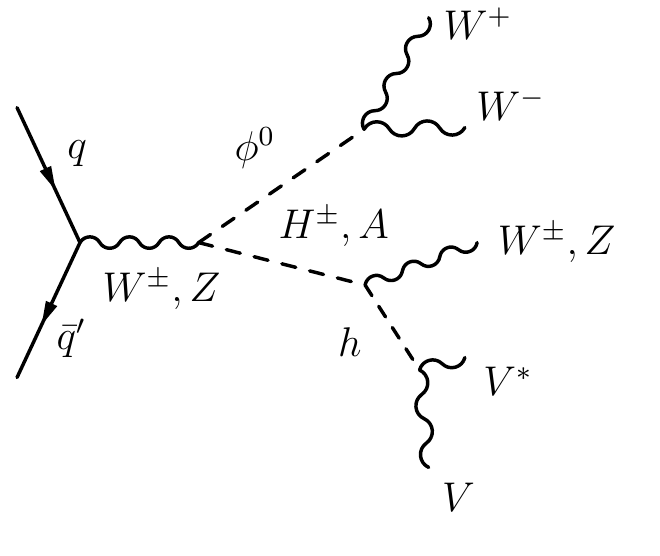}
        \caption{}
        \label{3l_feyn_h}
    \end{subfigure}
    \begin{subfigure}[t]{0.48\textwidth}
    \centering
        \includegraphics[scale=0.9]{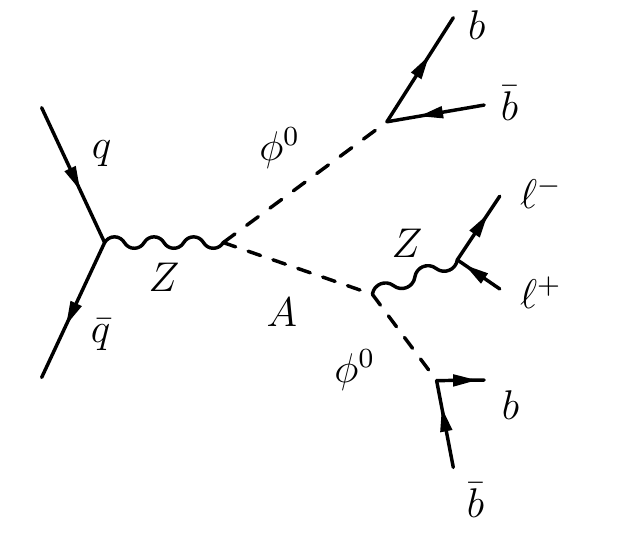}
        \caption{}
        \label{ossf3b_feyn}
    \end{subfigure}
    \begin{subfigure}[t]{0.48\textwidth}
    \centering
        \includegraphics[scale=0.9]{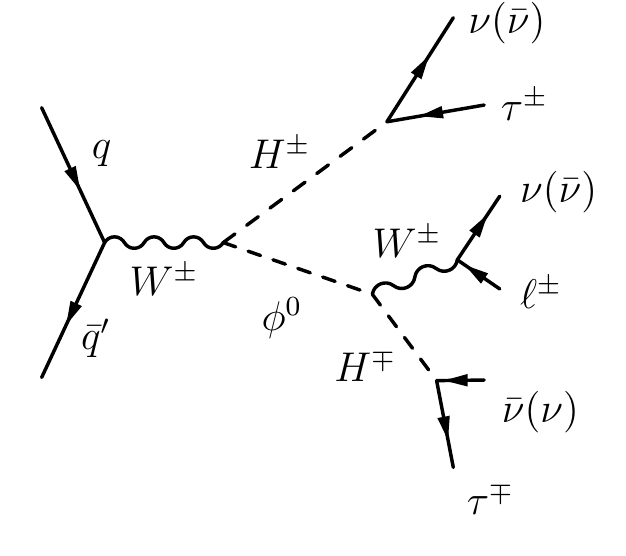}
        \caption{}
        \label{ssl0b_feyn}
    \end{subfigure}
    \caption{Examples of signal topologies that give rise to multi-lepton final states.}
\label{feynman_diagrams}
\end{figure*}

We have investigated a large number of processes in this model that may be possible to probe at the LHC.
The results of the investigation are summarized in \S{}\ref{sec:conclusions} (Tables~\ref{triangle_low}, \ref{triangle_up}, and \ref{diagonal}).
For optimistic benchmark models, there are many decay modes where a $5\si$ discovery is possible with $300$~fb$^{-1}$.
We will show below that there is significant additional parameter space
that can be probed by the high luminosity LHC ($3000$~fb$^{-1}$).
The most effective searches are multi-lepton channels, due to 
relatively low Standard Model backgrounds.
Illustrative event topologies leading to multi-lepton final states are
shown in Fig.~\ref{feynman_diagrams}.
Multilepton searches are standard parts of the LHC search program, so
this establishes that this model will be probed by new LHC data.
In addition, we identify one case where a novel search is sensitive,
involving a lepton pair (opposite sign, same flavor) plus 3 $b$ jets.

This paper is organized as follows.
In \S{}\ref{sec:model} we give additional details of our benchmark model
and its parameter space.
In \S{}\ref{sec:benchmarks} we give details of several benchmark studies. 
\S{}\ref{sec:conclusions} contains our conclusions, where we give projections of the search reach for both
300 fb$^{-1}$ and 3000 fb$^{-1}$ at the LHC.

%

\section{The Model}\label{sec:model}
We consider a model with 2 Higgs doublets $H_1, H_2$ with an approximate $\mathbb{Z}_2$
symmetry 
\[
H_1 \mapsto H_1, 
\qquad
H_2 \mapsto -H_2.
\]
In the $\ZZ$ symmetry limit, the Higgs potential is given by
\[
\eql{VZ2}
V_0 &= m_1^2 |H_1|^2
+ m_2^2 |H_2|^2 
+ \sfrac{1}{2}\la_1 |H_1|^4
+ \sfrac{1}{2}\la_2 |H_2|^4
\nonumber\\
&\qquad{}
+ \la_3 |H_1|^2 |H_2|^2
+ \la_4 |H_1^\dagger H_2|^2 
+ \sfrac{1}{2}\la_5 \bigl[ (H_1^\dagger H_2)^2 + \hc \bigr]
\]
All couplings can be chosen real by rephasing $H_{1,2}$, so the potential of model
naturally conserves $CP$ \cite{Gunion:2002zf}.
Note that the $\la_{3,4,5}$ terms can give unsuppressed mass splittings in the 
$H_2$ multiplet even in the $\mathbb{Z}_2$ symmetry limit.
We could even take the limit $m^2_2 \to 0$, in which case all of the 
mass of the exotic Higgs bosons comes from electroweak symmetry breaking. 
In particular, the term $|H_1|^2|H_2|^2$ contributes an electroweak-preserving
mass for $H_2$, which does not give rise to precision electroweak observables
such as $S$ and $T$.
The fact that this mass comes from electroweak breaking is instead reflected
in the fact that $H_2$ has large couplings to $H_1$. 
Such large Higgs couplings are therefore the 
smoking gun signal of this kind of non-decoupling electroweak symmetry 
breaking.
This particularly motivates the study of triple Higgs couplings in this class
of models.
We leave this study for future work.

We assume that $m_2^2 > 0$, so that in the $\mathbb{Z}_2$ symmetry limit
only  $H_1$ gets a VEV. We then have
\[\label{z2}
\begin{split}
v^ 2 & =  -\frac{2m_1^2}{\la_1} ,\\
m_h^2 &=\la_1v^2,\\
m_{\phi^0}^2 &=m^2_2+\frac{1}{2}(\la_3+\la_4+\la_5)v^2,\\
m_{A}^2 &=m^2_2+\frac{1}{2}(\la_3+\la_4-\la_5)v^2,\\
m^2_{H^{\pm}} &=m^2_2+\frac{1}{2}\la_3v^2,
\end{split}
\]
where $\phi^0,A, H^{\pm}$ are the physical fields that reside in $H_2$.
A big mass splitting in $m_A$ and $m_{H^{\pm}}$ violates custodial symmetry, which is severely constrained by electroweak precision tests. Therefore, from now on, we work in
the custodial symmetry limit $m_A=m_{H^{\pm}}$, which implies that $\la_4 =  \la_5$.

We also include $\mathcal{O}(\epsilon)$ terms that break $\ZZ$:
\[
\eql{VZ2break}
\begin{split}
\Delta V &= \Delta m^2(H_1^{\dagger}H_2 +\text{h.c.})
\\
&\qquad{}
+ \Delta \la  |H_1|^2(H_1^{\dagger}H_2 +\text{h.c.}) + \Delta \la ' |H_2|^2(H_1^{\dagger}H_2 +\text{h.c.})  
\end{split}
 \]
Not all of the couplings in \Eqs{VZ2} and \eq{VZ2break} are important for phenomenology.
This is because $\avg{H_2} =\mathcal{O}(\epsilon)$, 
and we are not interested in terms with more than 2 Higgs fields. 
The effects of $\la_2$ and $\De\la'$ are therefore suppressed by $\ep$,
and we can neglect them to get an overview of the phenomenology.
(We can think of $H_2$ as ``small.'')
Since we also set $\la_4 = \la_5$, we effectively have 7 parameters instead of 10: 
\[\label{parameters}
v,\ m^2_h,\ m^2_{\phi^0},\ m^2_A,\ \la_3,\ \Delta m^2,\ \Delta \la.
\]
The first two parameters are of course fixed by experiment to be
$m_h = 125$~GeV and $v = 246$~GeV, leaving 5 free parameters.
However, we will show that for small $\ep$ the phenomenology is
essentially determined by the mass spectrum of the new Higgs bosons.

Production of $\ZZ$ odd Higgs bosons comes from the couplings
such as $g_{Z A \phi^0}$, $g_{Z H^+ H^-}$, and $g_{W^+ H^- \phi^0}$,
which are fixed by gauge invariance.
Decays of heavier $\ZZ$ odd Higgs bosons to lighter $\ZZ$ odd Higgs bosons
are controlled by the same couplings.
The only additional couplings that we need are the ones that determine
the decay of the $\ZZ$ odd Higgs bosons to the $\ZZ$ even Higgs bosons
and Standard Model vector bosons.
For these we must consider the minimization of the Higgs potential.

We define the physical fields $h, \phi^0, A ,H^{\pm}$ in terms of the fields with the approximate the $\ZZ$ symmetry:
\begin{equation}
H_i=
\left (
  \begin{array}{c}
  H_i^+ \\
 \frac{1}{\sqrt{2}}( \tilde v_i+h_i+iA_i)
  \end{array}
\right ), 
\quad\quad i=1,2,\label{eq:z2basis}
\end{equation}
where $\tilde v_i, h_i, A_i, H_i^+$ are the VEV, CP-even neutral, CP-odd neutral and charged components in each doublet.
The physical pseudoscalar field is then given by
\[
A=A_2 +\epsilon_A A_1 +\mathcal{O}(\epsilon^2),
\]
with
\[
\eql{epA}
\epsilon_A=-\frac{\tilde v_2}{v} = O(\ep),
\]
where $\tilde v_2\equiv \left<H_2\right>$.
The physical scalars are
\begin{equation}
\left(
 \begin{array}{c}
h\\
\phi^0
  \end{array}
\right)=
\left(
 \begin{array}{cc}
1 & \ep_h\\
- \ep_h & 1\\
  \end{array}
\right)
\left(
 \begin{array}{c}
h_1\\
h_2
  \end{array}
\right) +\mathcal{O}(\ep^2),
\end{equation}
with
\[
\ep_h=\frac{1}{m_h^2-m_{\phi^0}^2}\Big{[}\frac{\tilde v_2}{v}(m_{\phi^0}^2-2m_{H^{\pm}}^2+\la_3 v^2)
+\Delta\la v^2\Big{]} = O(\ep).
\]
Using standard results from 2 Higgs doublet models,
together with $g_{h_1V V}\propto v_1$ and \Eq{epA}, we then obtain the interaction vertices
that control the decays of the lightest $\ZZ$ odd Higgs:
\begin{subequations}
\eql{hVV}
\[
& 
\ep_V \frac{m_Z}{v}(p_A+p_h)^\mu Z_\mu A h
\\
& i\ep_V \frac{m_W}{v}(p_{H^{\pm}}+p_h)^\mu W^\mp_\mu H^\pm h,
\\
& \ep_V \frac{m_V^2}{v} \phi (Z^\mu Z_\mu + 2 W^{+\mu} W^-_\mu)
\]
\end{subequations}
where
\[
\ep_V = \ep_A + \ep_h = O(\ep).\label{eq:ev}
\]
Here the 4-momenta are all defined to flow into the vertex.
We now discuss couplings of the $\ZZ$ odd Higgs bosons to fermions,
which are relevant for the decay of the lightest $\ZZ$ odd Higgs boson.
We define the fermions to be even under $\ZZ$, so in the $\ZZ$-symmetric limit, only Yukawa couplings involving $H_1$ are allowed.
This is a ``type I'' 2-Higgs doublet model, which naturally avoids non-Standard Model
flavor violation. 
When we include $\ZZ$ breaking, we must allow $O(\ep)$ Yukawa couplings
to $H_2$, so this model is no longer type I for $\ep \ne 0$.
We then have to worry about re-introducing
unacceptably large flavor violation at $O(\ep)$.
It may be interesting to consider the 
possibility that $\ep$ sufficiently suppresses non-Standard Model flavor violation.
Our focus is on direct searches for new Higgs bosons, so we 
will avoid flavor problems by making the phenomenological assumption 
that all flavor breaking is contained 
in a single set of Yukawa coupling matrices $y_u, y_d$ and $y_e$.
This is ``minimal flavor violation.'' 
Its validity depends on the UV completion of the theory having a single source of 
flavor breaking, at least to a very good approximation. 
With this assumption, the couplings of the Higgs fields to fermions is given by
\[
\begin{split}
\mathcal{L}_{\text{Yukawa}} &= {(y_u)}_{ij} \bar{Q}_{L_i}(H_1 + \ep_uH_2){u_R}_j 
+ {(y_d)}_{ij} \bar{Q}_{L_i}(H_1 + \ep_d H_2){d_R}_j \\
&\qquad{}
+ {(y_e)}_{ij}\bar{L}_{L_i}(H_1 + \ep_e H_2){e_R}_j + \text{h.c.}
\end{split}
\]
We will also make the phenomenological assumption that
\[
\ep_u \simeq \ep_d \simeq \ep_e.
\]
Then we have for any fermion $f$
\[
g_{\phi^0 ff} =g_{hff}(\ep_{u,d,e} -\ep_h).
\]
We see that the decays of $\phi^0, A$ and $H^{\pm}$ to fermions is controlled 
by the small parameter
\[
\ep_f \equiv \ep_{u,d,e} - \ep_h.
\]
It is natural to assume that $\ep_f \sim \ep_V$.
Note that both $\ep_V$ and $\ep_f$ involve $\ep_h$,
which depends on $\ZZ$ breaking in the Higgs potential. 
Therefore it is not natural to have $\ep_V \gg \ep_f$. 
If we have $\ep_f \gg \ep_V$, then fermion loops will induce $\ZZ$ breaking
in the Higgs potential.
For the top quark loop, we expect
\[
\Delta m^2 \gtrsim \ep_t \frac{3y_t^2}{8\pi^2}\La^2,
\]
where $\La$ is a UV cutoff. 
Even for $\La\sim\text{TeV}$ this is not suppressed. 

\begin{figure*}[t]
    \centering
    \begin{subfigure}[t]{0.5\textwidth}
    \centering
        \includegraphics[scale=0.7]{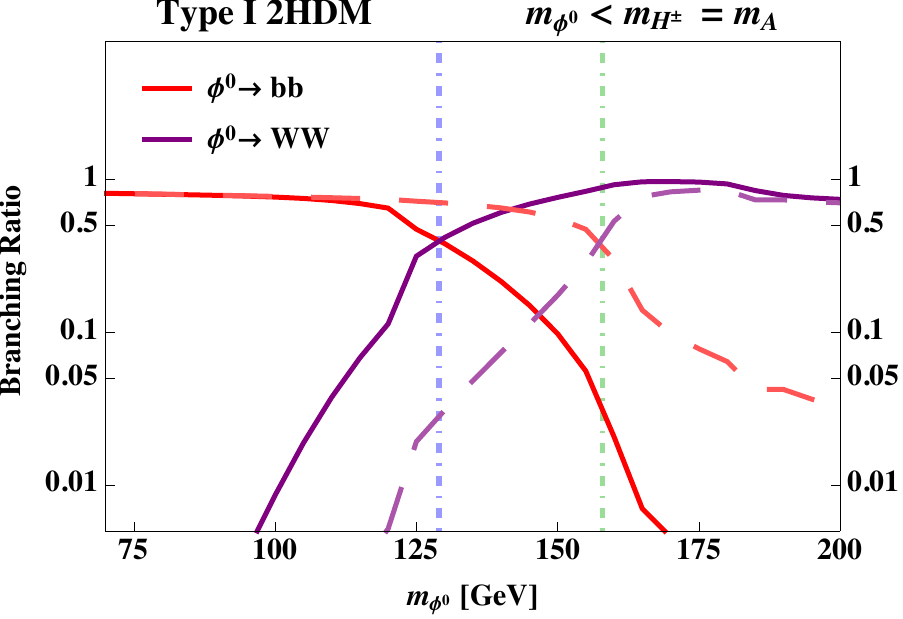}
        \label{HBR}
    \end{subfigure}
    \begin{subfigure}[t]{0.5\textwidth}
    \centering
        \includegraphics[scale=0.7]{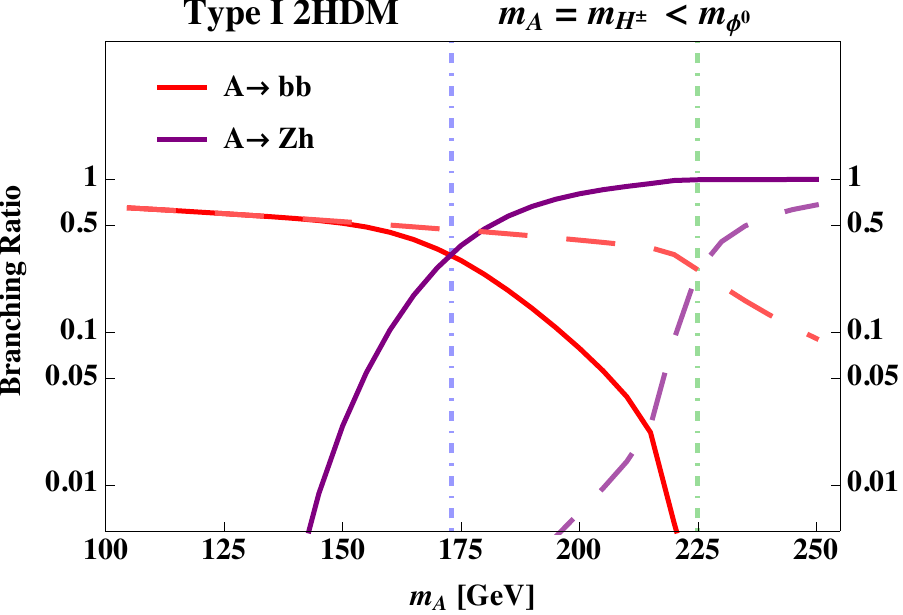}
        \label{ABR}
    \end{subfigure} 
    \begin{subfigure}[t]{0.5\textwidth}
    \centering
        \includegraphics[scale=0.7]{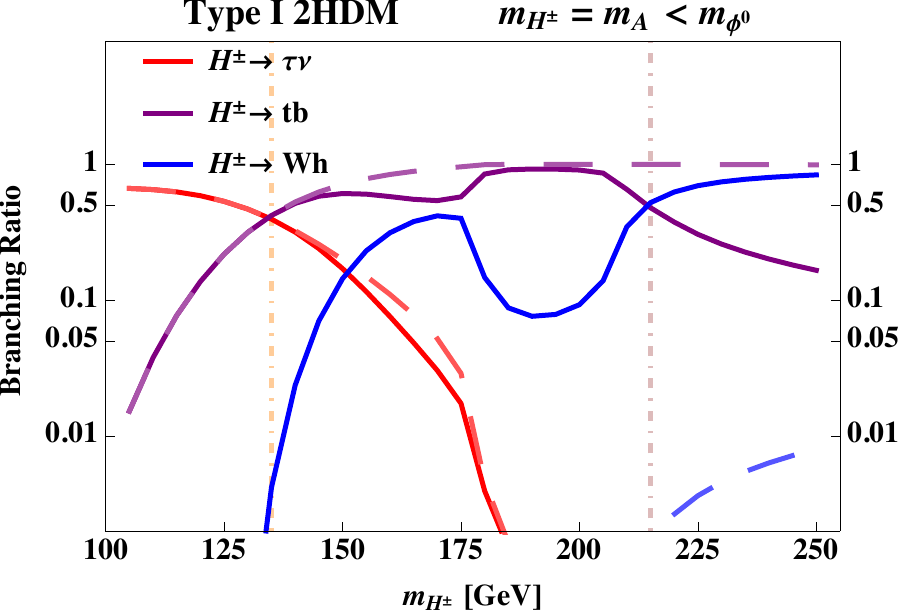}
        \label{CBR}
    \end{subfigure}
    \caption{Branching Ratios of the main decay modes of $\phi^0$, $A$ and $H^{\pm}$ to the SM particles. 
    The solid lines correspond to $r=1/5$ and the dashed lines are for $r=5$. }
\label{BR_degenerate}
\end{figure*}

Although we will assume $\ep_f \sim \ep_V$ in our study, 
the relative size of these suppressions is important for phenomenology 
because it determines the masses at which different decays become
dominant.
For example, if $\phi^0$ is the lightest $\ZZ$ odd Higgs boson,
it can decay either to $WW$ or $b\bar{b}$.
The decay to $b\bar{b}$ becomes dominant for $m_{\phi^0} \lsim 2m_W$,
but the precise mass for which this occurs is sensitive to the ratio
\[
r = \frac{\ep_f}{\ep_V}.
\]
Fig.~\ref{BR_degenerate} shows branching ratios of the main decay modes of $\phi^0$, 
$A$ and $H^{\pm}$ to the SM particles for $r=1/5$ and the dashed lines assume that $r=5$. 
The phenomenology therefore depends on this parameter in addition
to the spectrum of $\ZZ$ odd Higgs bosons.
This parameter affects only the reach of a given search,
so searches can be optimized only on the basis of the spectrum of
masses of the exotic particles.

For all the benchmark models considered in our paper, we found parameters
in the 2-Higgs doublet model parameter space that give an
experimentally acceptable contribution to the $S$ parameter.
This is easily accomplished despite the fact that the additional
Higgs bosons are light because they are approximately inert.
In addition, these models easily satisfy all perturbativity
constraints on the potential because the additional Higgs bosons
are all light.\footnote{After fixing the values of $\epsilon$ and the masses of the additional scalars, we still have $\Delta m^2$ as a free parameter in the scalar potential, that can be adjusted in such a way that unitarity, perturbativity and stability of the potential is assured.}

%


\section{Benchmark Studies} \label{sec:benchmarks}
In this section we study several benchmark models
with multi-lepton signals.
Simulated events for both signal and Standard Model backgrounds were generated by 
{\tt MadGraph5} \cite{Alwall:2014hca}, with showering and hadronization 
simulated by {\tt Pythia8} \cite{Sjostrand:2014zea},
and the detector response simulated by {\tt Delphes3} \cite{deFavereau:2013fsa}.
The leading order cross-sections of the signal and Standard Model
backgrounds for each channel are calculated by {\tt MadGraph5}. Several of the Standard Model backgrounds, such as $t\bar{t}$ and $W/Z+$jets have large NLO contributions, therefore we scale the LO cross sections of these processes with their corresponding K-factors\cite{Campbell:2011bn}.
Since we focus on the final states that contain leptons and $b$ jets, 
common selection requirements are applied to reconstructed jets, muons and electrons, before further selection requirements, optimized for each final
state, are applied.
Leptons are required to have a transverse momentum $\pt > 10~\GeV$ and pseudorapidity
$|\eta|<2.5$. We further require isolated leptons, as determined from
the isolation ratio $R_{\rm iso} = p_{Tj}/p_{T\ell}$
where $p_{Tj}$ is the clustered transverse energy, contained in
a cone of radius $\Delta R$ around the lepton, and $p_{T\ell}$
is the lepton transverse energy.  The lepton isolation requirement
used in this analysis is $\Delta R<0.2$ with $R_{\rm iso}<0.09$. 
Similar isolation criteria have been used by ATLAS for their
multilepton searches in LHC Run II\cite{Aaboud:2017dmy}.

Jets are required to satisfy $\pt > 20~\GeV$ and $|\eta| < 5$. The
$b$-tagging efficiency is taken to be the same as the default setting in {\tt Delphes3}. The remaining event selection is optimized for each individual channel, as described below.


\subsection{3 leptons off $Z\,$Peak}\label{3lep}
In the case when $\phi^0, H^{\pm}$ and $A$ are all relatively heavy,
they dominantly decay to final states that contain $W$ or $Z$. 
In particular, $H^{\pm}(A)$ can decay to $W(Z)\phi^0$ or $W(Z)h$ depending 
on the mass splitting between $\phi^0$ and $H^{\pm}(A)$.
In this scenario, pair-produced non-Standard Model $H$'s can decay to five to six on- or off-shell vector bosons (Figure \ref{3l_feyn}, \ref{3l_feyn_h}), therefore easily producing multiple leptons in the final state.

Asking for 3 light leptons has the advantage of a relatively low Standard Model background at LHC. Furthermore, given that pair produced $\phi^0H^{\pm}$, $\phi^0A$, $AH^{\pm}$ and $H^{+}H^{-}$ may all contain 3 leptons in their final states, this channel also benefits from high signal multiplicities. Its drawback is that signal decays cannot be reconstructed, hence the signal kinematic features are not prominent enough to discriminate them against SM backgrounds. As a result, this channel basically becomes a lepton counting channel, which can be potentially covered by the 3- lepton bin of general multi-lepton searches from ATLAS and CMS. 

\begin{figure*}[t!]
    \begin{subfigure}[t]{\textwidth}
    \centering
        \includegraphics[scale=0.8]{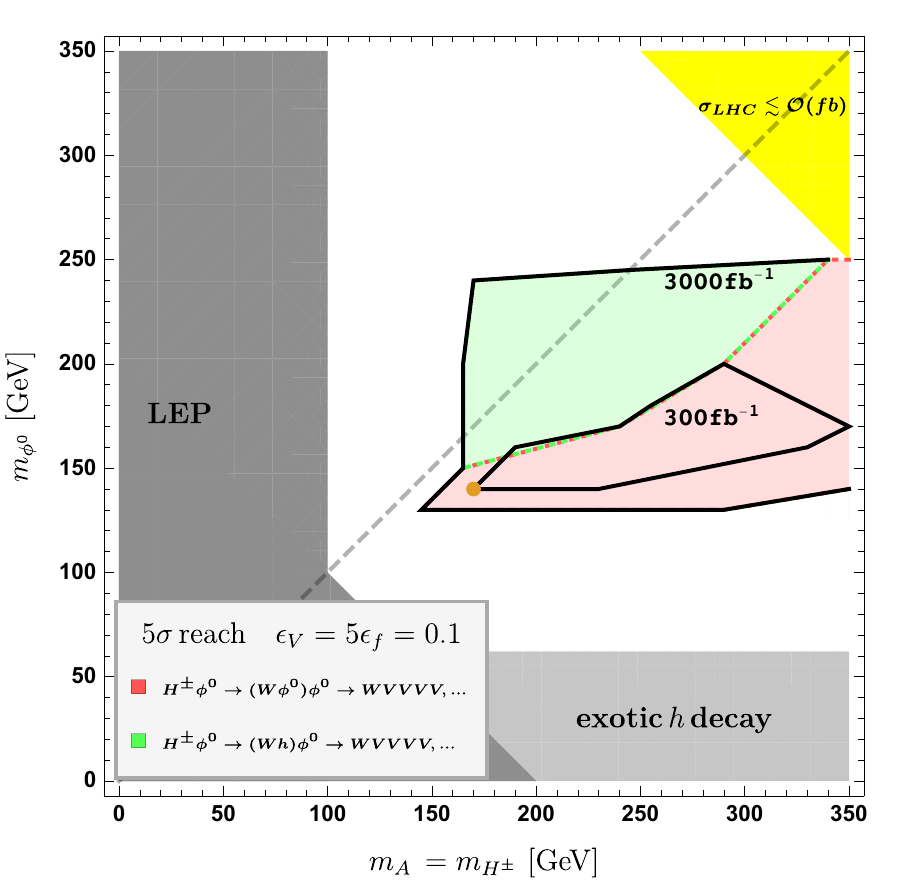}
        \label{theta01}
    \end{subfigure}
    \begin{subfigure}[t]{0.5\textwidth}
        \includegraphics[scale=0.8]{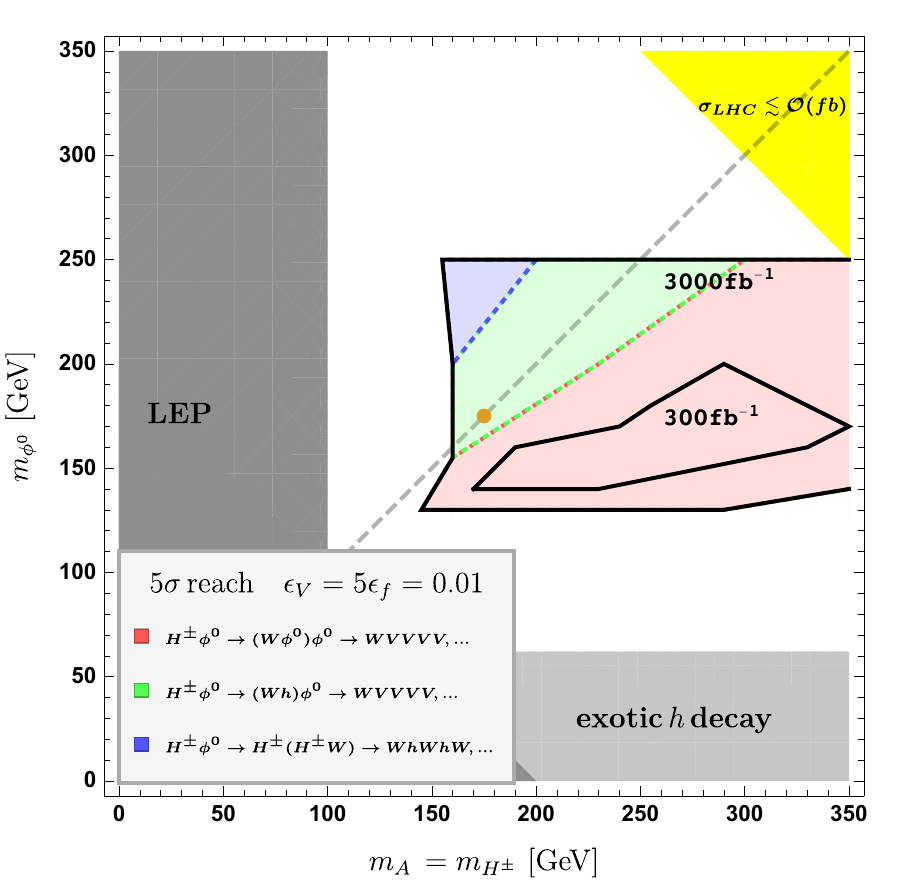}
        \label{theta001}
    \end{subfigure}
    \begin{subfigure}[t]{0.5\textwidth}
        \includegraphics[scale=0.8]{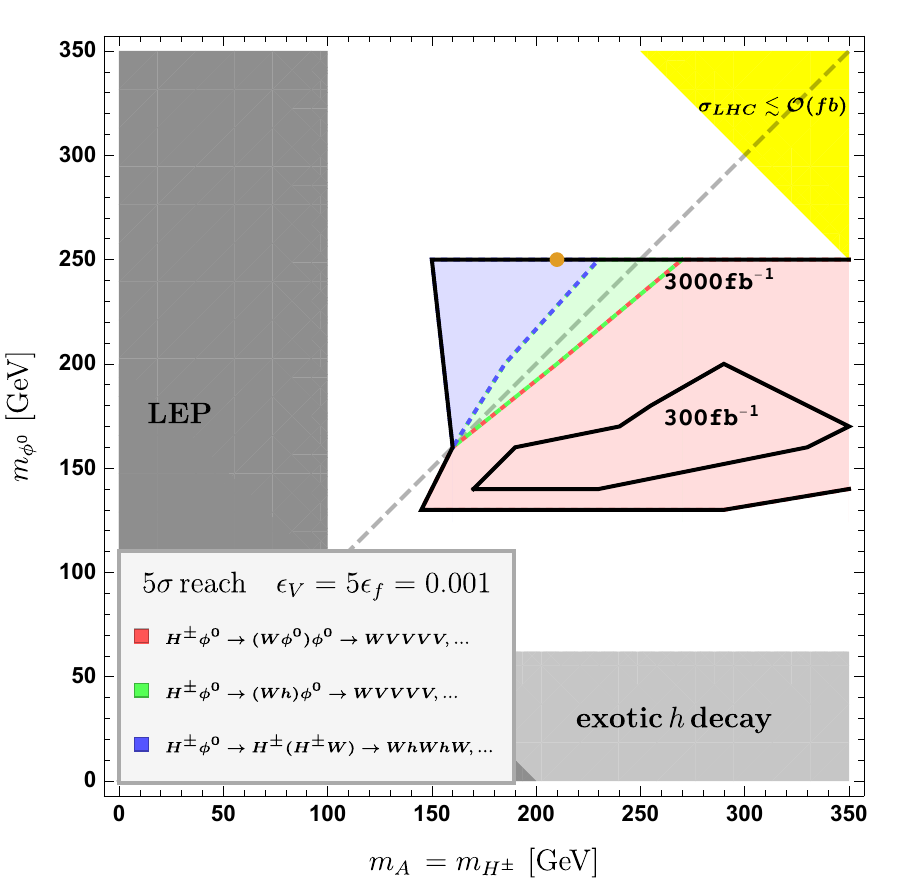}
        \label{theta0001}
    \end{subfigure} 
    \caption{Five $\si$ limits of the 3$\ell$ off $Z$ channel for different values of $\ep_V$ at LHC 13 TeV. The signal region is defined by $b$-veto, $\met >40$ GeV, $H_T > 300$ GeV and $N_j>2$. The $3000$ fb$^{-1}$ limit is further divided into three subregions where more than half of the signals come from each of the `colored' decays. The three benchmaks described in the text are marked by orange dots.}
\label{vary_theta}
\end{figure*}

Figure \ref{vary_theta} shows the main result of this search, where we draw the 5$\si$ contours reached at LHC run II and high-luminosity (HL) LHC. As we shall see, the overall 5$\si$ reaches are not affected by varying $\ep$s as long as $\ep_f\ll\ep_V$. The reason is that for any values of the $\ep$s considered, the final states of the exotic $H$'s decays always include combinations of the SM vector bosons and 125 GeV Higgs.
In Figure (\ref{decay}), we showed that the dominant decays for $A,\phi^0, H^{\pm}$ with smaller masses are to SM fermions, therefore they do not contribute to the multi-lepton signal.

\begin{table}
\centering
\begin{adjustbox}{max width=\textwidth}
\begin{tabular}{| l | l | l | l | l | }
\hline
Benchmark 1& $(m_{\phi^0},m_{A,H^{\pm}})=(140, 170)$ & $\ep_V=5\ep_f=0.1$ &$\mathcal{BR}_{\phi^0\to WW^*}\simeq 65\% $ & $\mathcal{BR}_{H^{\pm}(A)\to W^*(Z^*)\phi^0}\simeq 100\%$\\
&&&&\\
Benchmark 2& $(m_{\phi^0},m_{A,H^{\pm}})=(175, 175)$ & $\ep_V=5\ep_f=0.1$ &$\mathcal{BR}_{\phi^0\to WW^*}\simeq 100\% $ & $\mathcal{BR}_{H^{\pm}(A)\to W^*(Z^*)h}\simeq 40\%$\\
&&&&\\
Benchmark 3& $(m_{\phi^0},m_{A,H^{\pm}})=(250, 210)$ & $\ep_V=5\ep_f=0.001$ &$\mathcal{BR}_{\phi^0\to AZ^*\to(hZ)Z^*}\simeq 15\% $ & $\mathcal{BR}_{\phi^0\to H^{\pm}W^*\to(hW)W^*}\simeq 30\%$\\
\hline
\end{tabular}
\end{adjustbox}
\caption{Details of the benchmarks for the 3 leptons off $Z\,$peak search.}\label{tablebench1}
\label{benchmarks}
\end{table}

Table \ref{benchmarks} lists three benchmarks that are representative of each type of decay based on the assumptions on the mass hierarchy of $\phi^0, A$ and $H^{\pm}$. 
Benchmark 1 (B1) gives an example of the scenario in which $m_{\phi^0}<m_{H^{\pm},A}$ and the mass splitting between $\phi^0$ and $H^{\pm}(A)$ is sufficient to allow $H^{\pm}(A)$ to decay to $W^\pm(Z)\phi^0$. This corresponds to the region below the diagonal line in Figure \ref{vary_theta}.
Moving towards the diagonal, the mass splitting shrinks and the dominating decay modes of $H^{\pm}(A)$ are through $W^\pm(Z)h$, as long as $\ep_V>\ep_f$. Benchmark 2 (B2) corresponds to this region.
Finally, as we cross the diagonal, where $m_{\phi^0}>m_{H^{\pm},A}$, $\phi^0\to H^{\pm}W^*$ or $AZ^*$ can take over $\phi^0\to VV$, provided $\ep_V$ is very small ($\lsim 10^{-2}$). Even though $\phi^0$'s decay to $A(H^{\pm})V^*$ is not kinematically favorable, it is not suppressed by $\ep_V$. Benchmark 3 (B3) corresponds to this scenario.

The main Standard Model backgrounds include dibosons, $t\bar{t}V$, and $t\bar{t}$ or $Z$ plus jets with one fake/non-prompt (FNP) lepton.
To estimate the FNP leptons, we simulate $Z$ plus jets and $t\bar{t}$, both of which are then decayed to include at least two leptons.
Then, we select events that contain at least two reconstructed leptons and one jet, assuming a flat jet-faking-lepton rate.
We match their contributions to the $3\ell$ bin in Figure 2(d) of the 36 $\text{fb}^{-1}$ ATLAS multi-lepton search \cite{Aaboud:2017dmy} and extract the jet-faking-lepton rate $\sim 8 \times 10^{-4}$.

For the preselections, we require a $b$-veto, at least 3 leptons with $\pt$ of the leading (sub-leading) lepton $>20\, (15)$ GeV. If a pair of OSSF leptons are found, we require that their invariant mass $\notin (m_Z-15,m_Z+15)$ GeV. Since the signals are relatively massive and typically consist of five or six vector bosons, with more than half of them undergoing leptonic decays, we also require the missing energy  $\met>40$ GeV and $H_T > 300$ GeV, where $H_T$ is the scalar sum of the lepton and jet $\pt$'s.

Due to the limited number of signal events and their lack of prominent kinematic features, we are only able to place a final cut on the number of jets, $N_j$.
Table \ref{signal170_140} gives the signal yields for the three benchmarks and the SM backgrounds, assuming an integrated luminosity of $300\mbox{ fb}^{-1}$.
The signal corresponding to B1 is the only one that reach a $5\sigma$ significance for $300\mbox{ fb}^{-1}$.\footnote{For the significance $Z$, we use the expression \cite{Cowan:2010js}: $Z = \sqrt{2 \left[(S+B)\times \ln\left(1+S/B\right)-S \right]}$.} For B2 (B3), approximately $1700\mbox{ fb}^{-1}$ ($3000\mbox{ fb}^{-1}$) is required to achieve a 5$\si$ significance. Compared to B1, B2 and B3 perform much worse, mainly because $ h\to VV$  is not the dominant decay mode for a 125 GeV Higgs.

From the benchmark studies, it can be seen that in the case of a small $r(\equiv\ep_f/\ep_V$) and large Higgs masses, $\phi^0$, $A$ and $H^{\pm}$ dominantly decay to $V+X$. Regardless of what assumptions are made about their mass hierarchy, the pair produced exotic $H$s can always contribute to the signal $3\ell$ off $Z$. 
We also investigate whether our results will be affected by varying the absolute values of $\ep$s.
In Fig.~\ref{vary_theta}, the $5\si$ contours are plotted for three different values of $\ep_V$ with $r$ held fixed. 
As the $\epsilon$'s become smaller and smaller, the suppression due to $\epsilon^2$ in the $\ZZ$-odd $\phi^0$ decaying to SM fields becomes comparable to the phase space suppression of $\phi^0\to H^{\pm}W^{\mp}$, and the latter starts to contribute to the signal region. Therefore, one sees a slight increase in the reach of the search as the $\epsilon$'s decrease.
Despite that the dominant decays of $\phi^0$, $H^{\pm}$ and $A$ can be different under the variation of $\ep_V$, they all end up contributing to the signals that we are looking for.
As a result, the 5$\si$ limit contour does not depend much on the absolute values of $\ep_V$ or $\ep_f$. As long as $\ep_V$ is much larger than $\ep_f$, the three types of decays compliment each other.
\begin{table}
\centering
\begin{adjustbox}{max width=\textwidth}
\begin{tabular}{ l  c  c  c  c }
\hline
& $\si$(fb) & initial@300fb$^{-1}$ &pre-selection &  final selection\\
\hline
\multirow{2}{*}{$\ZZ$ odd Higgs $(m_{H^{\pm},A},m_{\phi^0})$} & & & & \\
& & & &\\
\hline
Benchmark 1: $(170,140), \ep_V=5\ep_f=0.1$& & & &\\
\hline
 $\phi^0H^{\pm}\to\phi^0({W^{\pm}}^*\phi^0),\phi^0\to WW^*$ &25&$7478$ &41 & 23\\

 $\phi^0A\to\phi^0(Z^*\phi^0),\phi^0\to WW^*$ &14&$4056$  &19 & 14\\

 $H^{\pm}A\to({W^{\pm}}^*\phi^0)(Z^*\phi^0),\phi^0\to WW^*$ &15&$4310$  &31 & 23\\

 $H^+H^-\to({W^+}^*\phi^0)({W^-}^*\phi^0),\phi^0\to WW^*$ &9&$2535$  &24 & 18\\
 \hline
B1 Total &  &  &  &  78\\
\hline
\hline
Benchmark 2: $(175,175), \ep_V=5\ep_f=0.1$& & & & \\
\hline
 $\phi^0H^{\pm}\to(W^+W^-)({W^{\pm}}^*h)$ &18&$5400$ &15 &9 \\

 $\phi^0A\to(W^+W^-)(Z^*h)$ &10&$3000$ &21&  17\\

 $H^{\pm}A\to({W^{\pm}}^*h)(Z^*h)$ &7&$2100$ &5 &6  \\

 \hline
B2 Total &  & &   & 32 \\
\hline
\hline
Benchmark 3: $(210,250), \ep_V=5\ep_f=0.001$& & & & \\
\hline
 $\phi^0H^{\pm}\to(V^*H^{\pm}/A)({W^{\pm}}^*h)$ &5&$1500$ &7 &6 \\

 $\phi^0A\to(V^*H^{\pm}/A)(Z^*h)$ &8&$2400$ &7&7\\

 $H^{\pm}A\to({W^{\pm}}^*h)(Z^*h)$ &7&$2100$ & 12&11 \\

\hline
B3 Total &  &  & &24  \\
\hline
\hline
\multirow{2}{*}{Standard Model backgrounds:} & & &  &\\
& & & &\\
\hline

$W^{\pm}Z\to(\ell^{\pm}\nu)(\ell^+\ell^-)$ &$1300$&$3.9\times10^5$ &190 & 44 \\

$ZZ,\,Z$\textrightarrow$\ell^+\ell^-$ &124&$3.7\times10^4$ &24 & 9 \\

$t\bar{t}V$ &900& $2.7\times10^5$ & 99 & 39\\

$VVV$ &440& $1.3\times10^5$ & 65 & 8\\

$hW,\, W\to\ell\nu$ &6& $1.8\times10^3$ & 13 & 3\\

di-leptonic $t\bar{t}$ (FNP) &$7.8 \times10^4$&$2.3 \times10^7$&196 & 95\\

di-leptonic $tWj$(FNP) &$0.5\times10^4$ &$1.5\times10^6$ &21 & 6\\

$Z+$jets, $Z$\textrightarrow$\ell^+\ell^-$ (FNP) &$2.3\times10^6$ &$6.9\times10^8$  &85 &13\\

di-leptonic $WW$ (FNP) &$1.0 \times10^4$&$2.9 \times10^6$&22 & 3\\

\hline
SM Total &  &  &   &220\\
\hline
\end{tabular}
\end{adjustbox}
\caption{Signal and the background yields for the channel 3$\ell$ off $Z$, assuming an integrated luminosity of $300\, \text{fb}^{-1}$. To estimate the number of events with FNP leptons, a flat jet-faking-lepton rate of $8\times10^{-4}$ is used. The preselections are 3$\ell$ off $Z$, $b$-veto, $\met >40$ GeV, $H_T > 300$ GeV and the final selection is $N_j>2$.}
\label{signal170_140}
\end{table}
%


\subsection{OSSF leptons with 3 $b$ jets}\label{ossf_3b}
The 3$\ell$ off $Z$ search above targets the parameter space with relatively massive $\ZZ$ odd Higgs particles. In this section, we look at a relatively light $\phi^0$ ($\lesssim 120$ GeV), where $\phi^0\to b\bar{b}$ becomes the dominant decay mode.

If $(m_{H^{\pm}}=)m_A >m_{\phi^0}$, $A$ predominantly decays to $\phi^0Z^{(*)}$. One interesting channel to consider is depicted in Figure \ref{ossf3b_feyn}, where $pp\to\phi^0A\to\phi^0(\phi^0Z^{(*)})\to(b\bar{b})(b\bar{b}\ell^+\ell^-)$ gives a final state that consists of a pair of opposite-sign same-flavor (OSSF) leptons and four $b$s.
Therefore, we ask for a pair of OSSF leptons with the leading (sub-leading) lepton $\pt > 20 (15)$ GeV, and at least 4 jets with 3 $b$-tagged jets. Since there is no invisible particles for the signal process, we also require $\met<50$ GeV as part of the preselections.

The dominating SM backgrounds are $Z+$jets, di-leptonic $t\bar{t}$ and single top production. Other SM backgrounds include di-bosons, $Vh$ and fake/non-prompt leptons, but they are negligible compared to the first three SM processes \cite{Aaboud:2019nan}.

Depending on whether the mass difference between $A$ and $\phi^0$ is greater than 91 GeV or not, this channel is further divided into the on- and off-shell $Z$ signal regions. Below we give detailed benchmark studies focusing on each region. For both choice of benchmarks, we further assume that $r\equiv \ep_f/\ep_V=5,\ep_f=0.1$.
Under these assumptions, $\mathcal{BR}_{\phi^0\to b\bar{b}}$ is approximately 80\% and $\mathcal{BR}_{A\to Z\phi^0}$ almost 100\%.


\subsubsection{off $Z: (m_A,m_{\phi^0}) = (150,70)$ GeV}\label{ossf_3b_offz}

\begin{figure*}[t!]
      \begin{subfigure}[t]{0.5\textwidth}
        \includegraphics[scale=0.4]{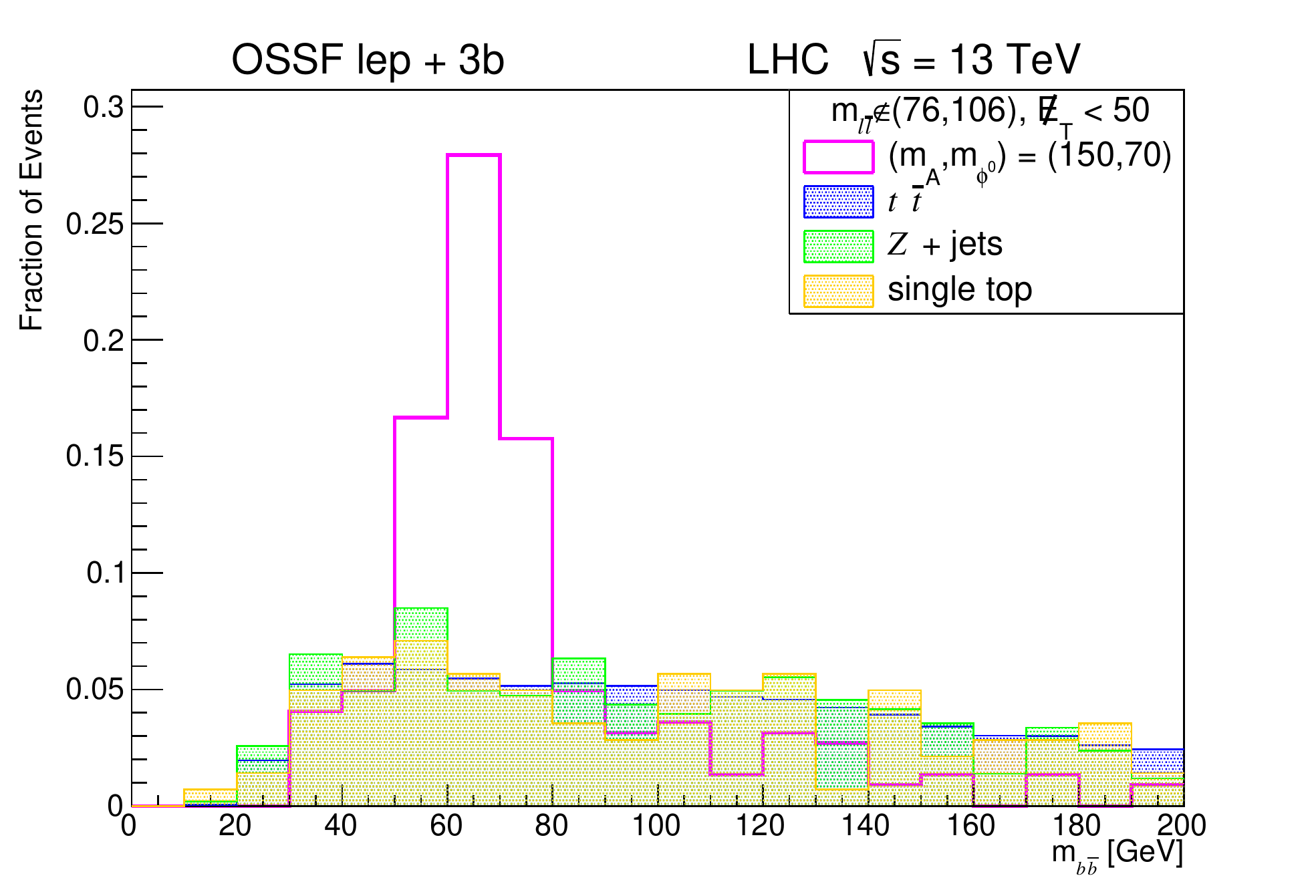}
        \caption{}
    \end{subfigure} 
    \begin{subfigure}[t]{0.5\textwidth}
        \includegraphics[scale=0.4]{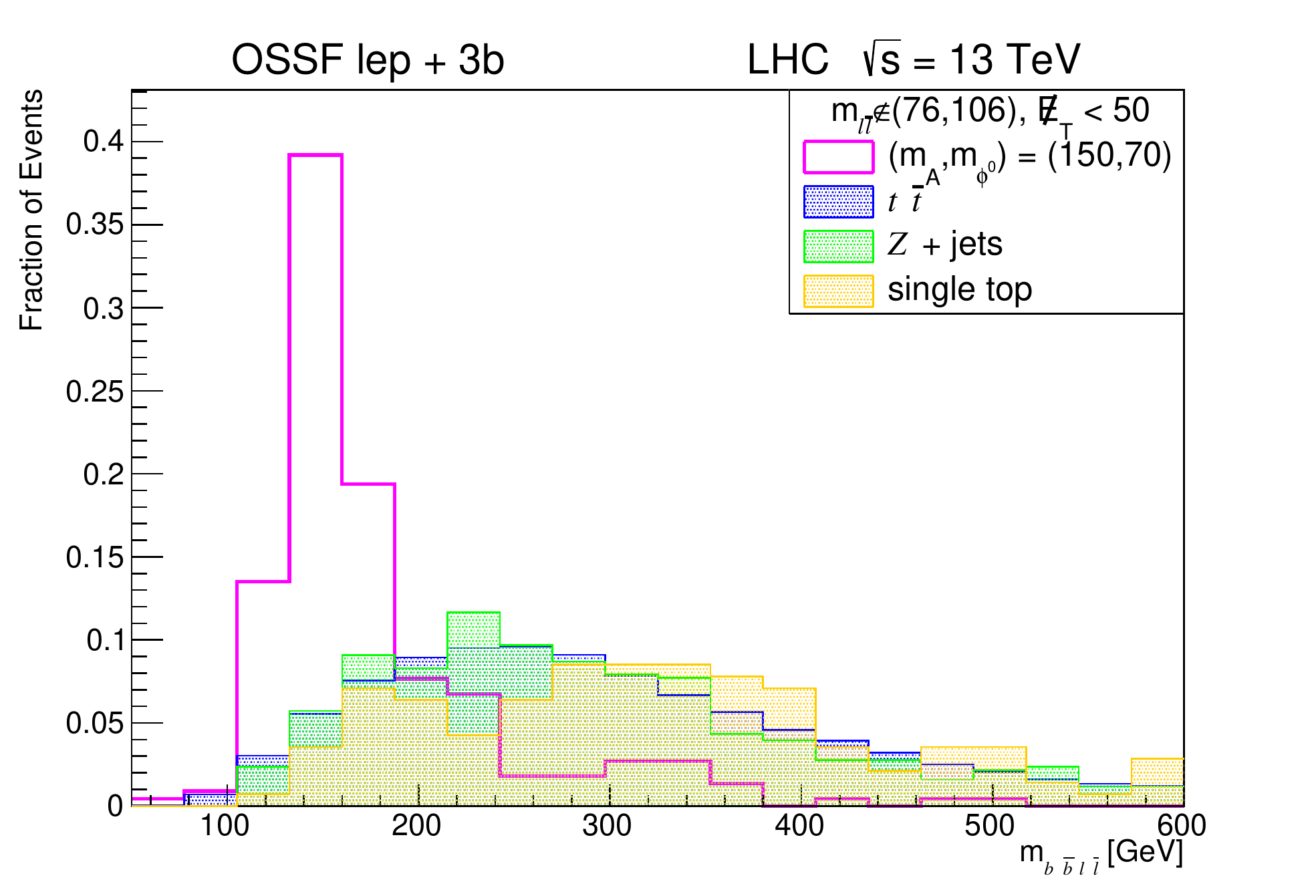}
        \caption{}
    \end{subfigure}
       \begin{subfigure}[t]{0.5\textwidth}
        \includegraphics[scale=0.4]{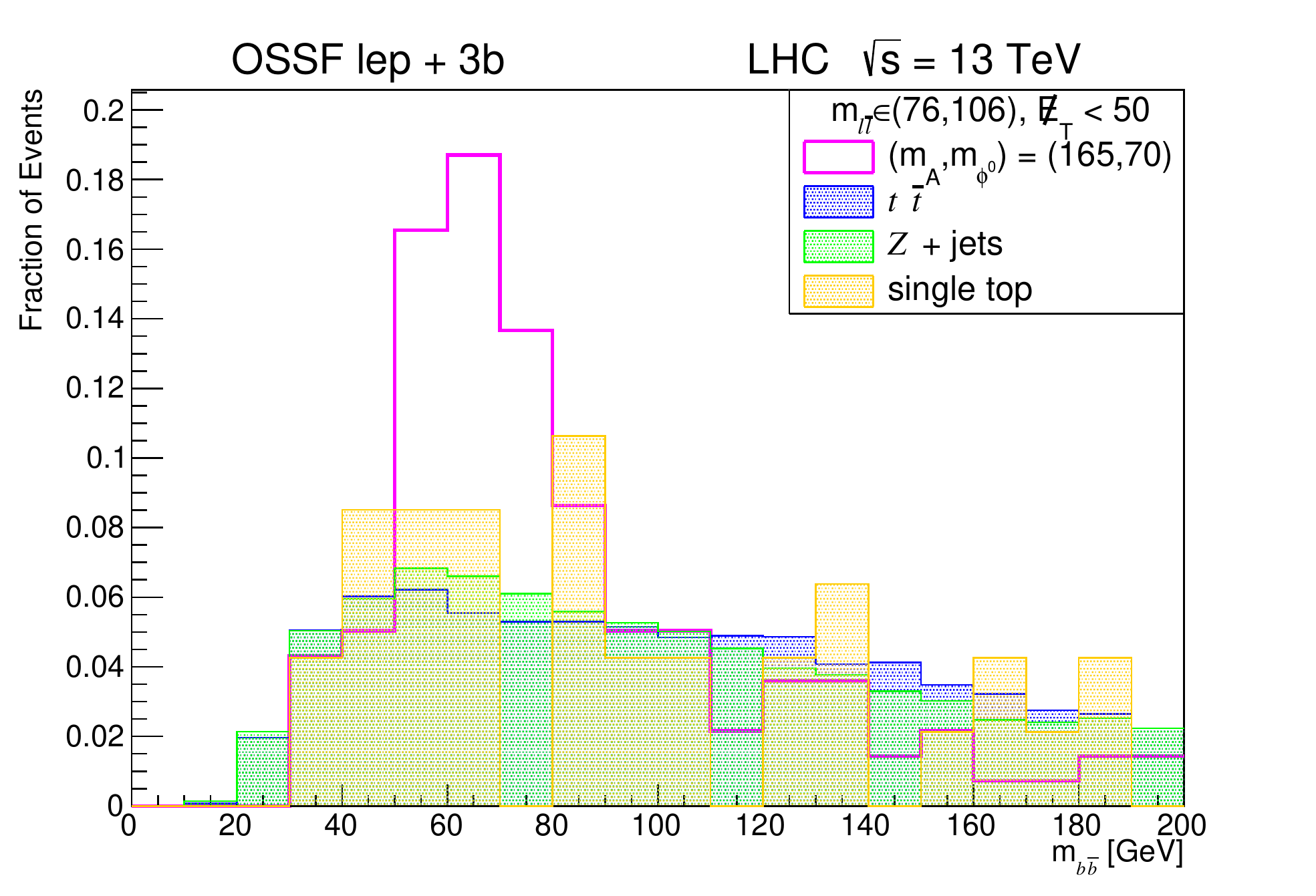}
        \caption{}
    \end{subfigure} 
    \begin{subfigure}[t]{0.5\textwidth}
        \includegraphics[scale=0.4]{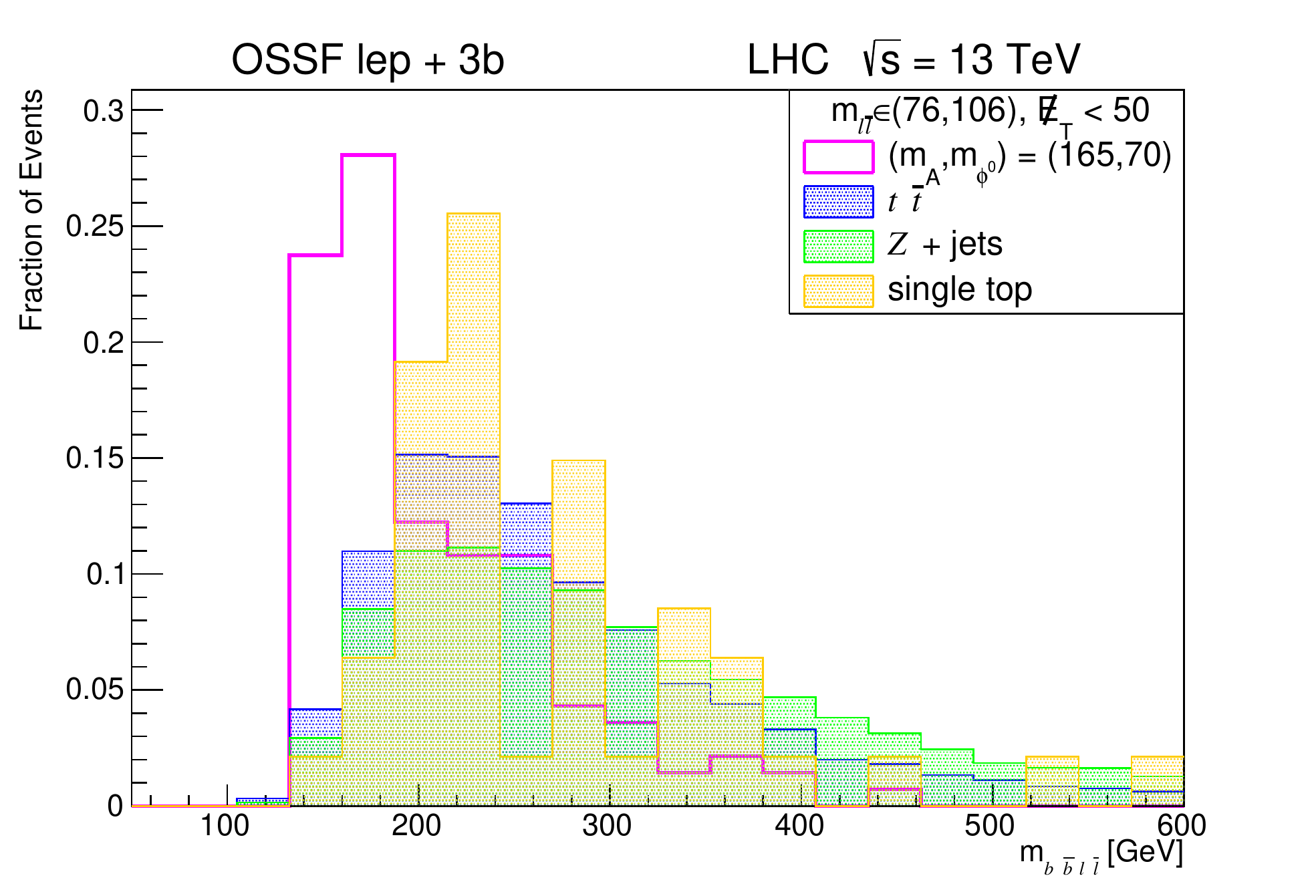}
        \caption{}
    \end{subfigure}

    \caption{Distributions of $m_{b\bar{b}}$ and $m_{\ell^+\ell^-bb}$ for the OSSF leptons and $3b$ signal (Sec.\ref{ossf_3b}) and the main SM backgrounds after pre-selections. Both signal benchmarks have $r =5,\ep_f=0.1$. The preselections are OSSF $\ell$ pair off (on) $Z$, $N_j>3$ with at least 3 $b$-tagged, $\met<50$ GeV.}
\label{ossf3b_offz}
\end{figure*}

After applying the preselections discussed above, we try to reconstruct the entire decay chain for the signal.
Since both $\phi^0$s decay to $b\bar{b}$, we assume that the jet with the highest transverse momentum out of the non-$b$-tagged jets to be the fourth $b$. To reconstruct the $\phi^0$s, 
we choose the combination of the jets that minimizes $(\Delta\phi_{j_1,j_2})^2+(\Delta\phi_{j_3,j_4})^2$. Since $A$ decays via $\phi^0$ and $Z^{(*)}$, we then reconstruct $A$ using the combination of the two leptons and the reconstructed $\phi^0$ that has a smaller value in $|\Delta\phi|$ . Figure \ref{ossf3b_offz} shows the reconstructed $A$ and $\phi^0$ mass distributions for signal and backgrounds after the preselections. As can be seen, both show prominent resonances for the signal, hence can be used to effectively suppress the backgrounds.
The final selections are $\met/\sqrt{H_T}<2$ GeV$^{1/2}$, $|m_{b\bar{b}}-m_{\phi^0}|<15$ GeV, $m_{b\bar{b}\ell\bar{\ell}}-m_{A}<20$ GeV.
Table \ref{signal70_150} gives the yields of the signal and the dominating backgrounds assuming an integrated luminosity of 300 $\text{fb}^{-1}$. To achieve a significance of 5$\si$, we need approximately 700 fb$^{-1}$. 

\begin{table}[t]
\centering
\begin{adjustbox}{max width=\textwidth}
\begin{tabular}{ l  c  c  c  c c c }
\hline
 & $\si$(fb) & initial@300fb$^{-1}$ &\multicolumn{2}{c}{pre-selection} & \multicolumn{2}{c}{final selection}\\
\hline
\hline
$\ZZ$ odd Higgs ($m_A,m_{\phi^0}$)& & & \multirow{2}{*}{B1}&\multirow{2}{*}{B2} & \multirow{2}{*}{B1} &\multirow{2}{*}{B2} \\
\scriptsize{$\phi^0A \to\phi^0(\phi^0Z)\to(b\bar{b})(b\bar{b}\ell^+\ell^-)$} & & & & & &\\
\hline
Benchmark 1: (150,70)&10& 3000 &67 &- &15&-\\
Benchmark 2: (165,70)& 12 &3600 &- &50 & -&12\\
\hline
\hline
SM Backgrounds & & &  &&&\\
\hline
di-leptonic $t\bar{t}$ & $78000$&$2.34\times10^7$ & 6554  &1634  &15&4 \\
di-leptonic $tW+$ jets &4800&$1.44\times10^6$& 136 & 45& 1&0\\

$Z b\bar{b}j,Z$\textrightarrow$\ell^+\ell^-$ &$103500$&$3.11\times10^7$ &185&3986&1&25\\
$Zbb\bar{b}\bar{b},Z$\textrightarrow$\ell^+\ell^-$ &980& $2.9\times10^5$ & 39&856 &0&2\\
\hline
SM Total & - & - & -  &-&17&31 \\
\hline
\end{tabular}
\end{adjustbox}
\caption{Signal and background yields for OSSF leptons plus 3 $b$s assuming an integrated luminosity of 300 fb$^{-1}$. The signal benchmarks both have $\ep_f=5\ep_V=0.1$. The preselections for Benchmark 1 (2) are OSSF $\ell$ pair off (on) $Z$, $N_j>3$ with at least 3 $b$ tagged and $\met<50$ GeV. The final selections for B1 and B2 are as described in the text above.}
\label{signal70_150}
\end{table}


\subsubsection{on $Z: (m_A,m_{\phi^0}) = (165,70)$ GeV}\label{ossf_3b_onz}

This benchmark produces an on-shell $Z$ in its decay, therefore we apply the same preselections as before, except for requiring an on-shell $Z$ instead of an off-shell $Z$. We repeat the analysis from Section \ref{ossf_3b_offz}.
The final selections are  $\met/\sqrt{H_T}<2$ GeV$^{1/2}$, $|m_{b\bar{b}}-m_{\phi^0}|<20$ GeV and $|m_{b\bar{b}\ell\bar{\ell}}-m_{A}|<10$ GeV.
Table \ref{signal70_150} gives the signal and background yields assuming an integrated luminosity of 300 $\text{fb}^{-1}$.
To achieve a significance of 5$\si$, we need roughly $1800\mbox{ fb}^{-1}$.
From the two benchmarks we studied, the on-shell $Z$ case performs much worse compared to the off-shell $Z$ case.


\subsection{2 Same-Sign leptons}\label{2ssl}

The search channel above targets a light $\phi^0$. In this subsection, we consider a light $H^{\pm} (A)$.
If $m_{\phi^0}>m_{H^{\pm}}(=m_A)$, $\phi^0\to H^{\pm}{W^{\mp}}^{(*)}$ or $AZ^*$ become the dominant decay. If $H^{\pm}$ is lighter than 130 GeV, it decays to $\tau\nu$ predominantly. As depicted in Fig.~\ref{ssl0b_feyn}, where $pp\to\phi^0H^{\pm}\to(H^{\pm}W^{\mp})H^{\pm}$ with $H^{\pm}\to \tau\nu$, if $W$ further decays leptonically, we can easily obtain a final state of $\ell^{\pm}\ell^{\pm}$ or $\tau_h^{\pm}\ell^{\pm}$, where $\ell$ represents $e$ or $\mu$ and $\tau_h$ a $\tau$-tagged jet.

For this search, we only consider the final states $\mu^{\pm}\mu^{\pm}$ or $\mu^{\pm}\tau_h^{\pm}$. 
$\tau_h^{\pm}\tau_h^{\pm}$ is not included because it suffers from a huge multi-jet background without light leptons. Electrons are not considered here because the charge misidentification is non-negligible for electrons.
The benchmark we choose to work with is $(m_{\phi^0},m_{H^{\pm}})=(160,110)$ GeV with $r=1/5,\epsilon_V=0.1$,
where $\mathcal{BR}_{\phi^0\to H^{\pm}W^{\mp}}$ is 80\% approximately and $\mathcal{BR}_{H^{\pm}\to\tau\nu}$ 65\% approximately.

The main irreducible backgrounds are dibosons, $t\bar{t}V$, $VVV$. The SM backgrounds with one fake/non-prompt (FNP) lepton or one fake $\tau_h$ come from $W$ or $Z$ plus jets and $t\bar{t}$. The fake rate is estimated to be approximately $10^{-4}$. 

For preselections, we ask for two same-sign muons or one muon plus one same-sign $\tau$-tagged jet. 
Events that have any $b$s are vetoed. We further require that $\met >85$ GeV, because the signal has multiple invisible particles in its final state.

\begin{figure*}[t!]
 \begin{subfigure}[t]{0.5\textwidth}
        \includegraphics[scale=0.4]{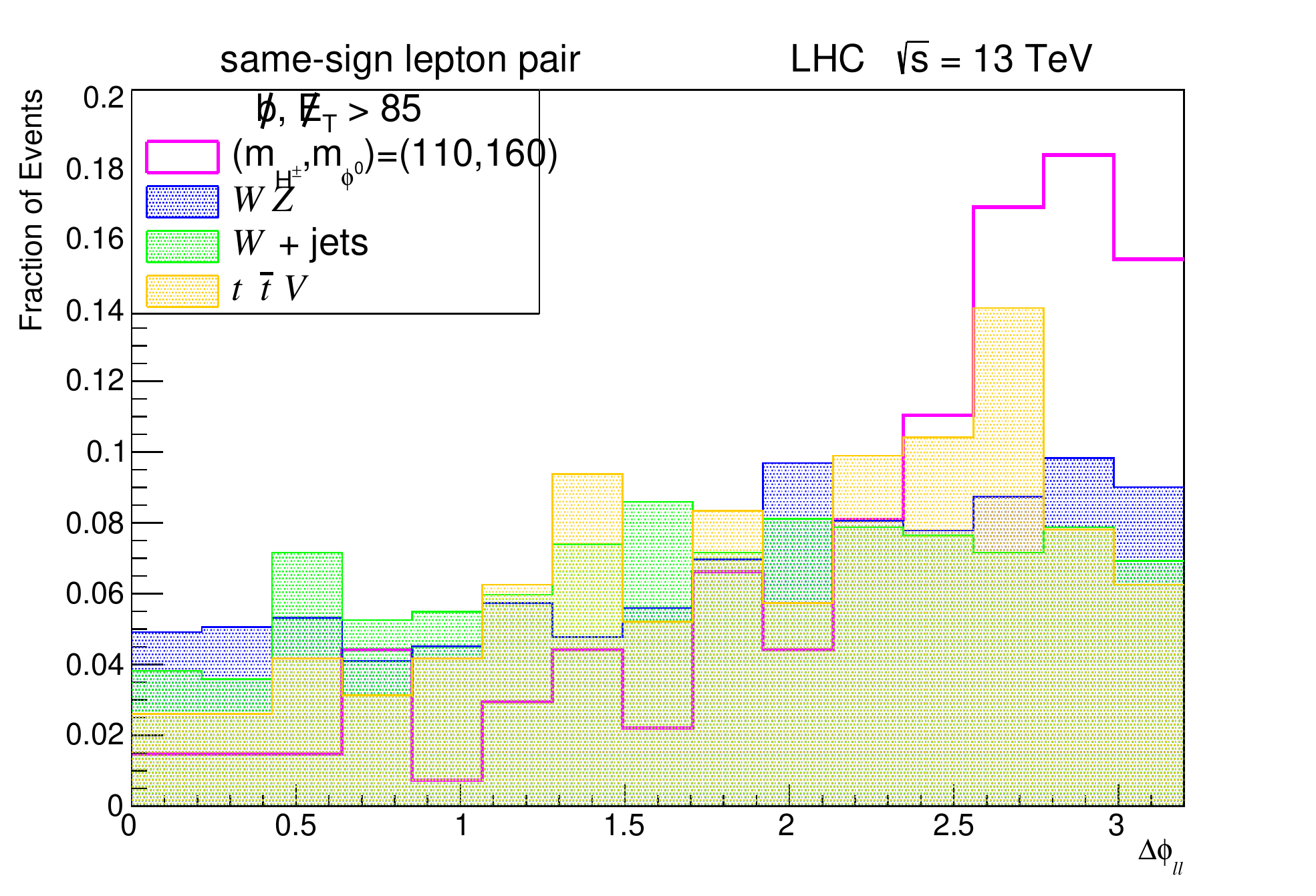}
        \caption{}
    \end{subfigure}
    \begin{subfigure}[t]{0.5\textwidth}
        \includegraphics[scale=0.4]{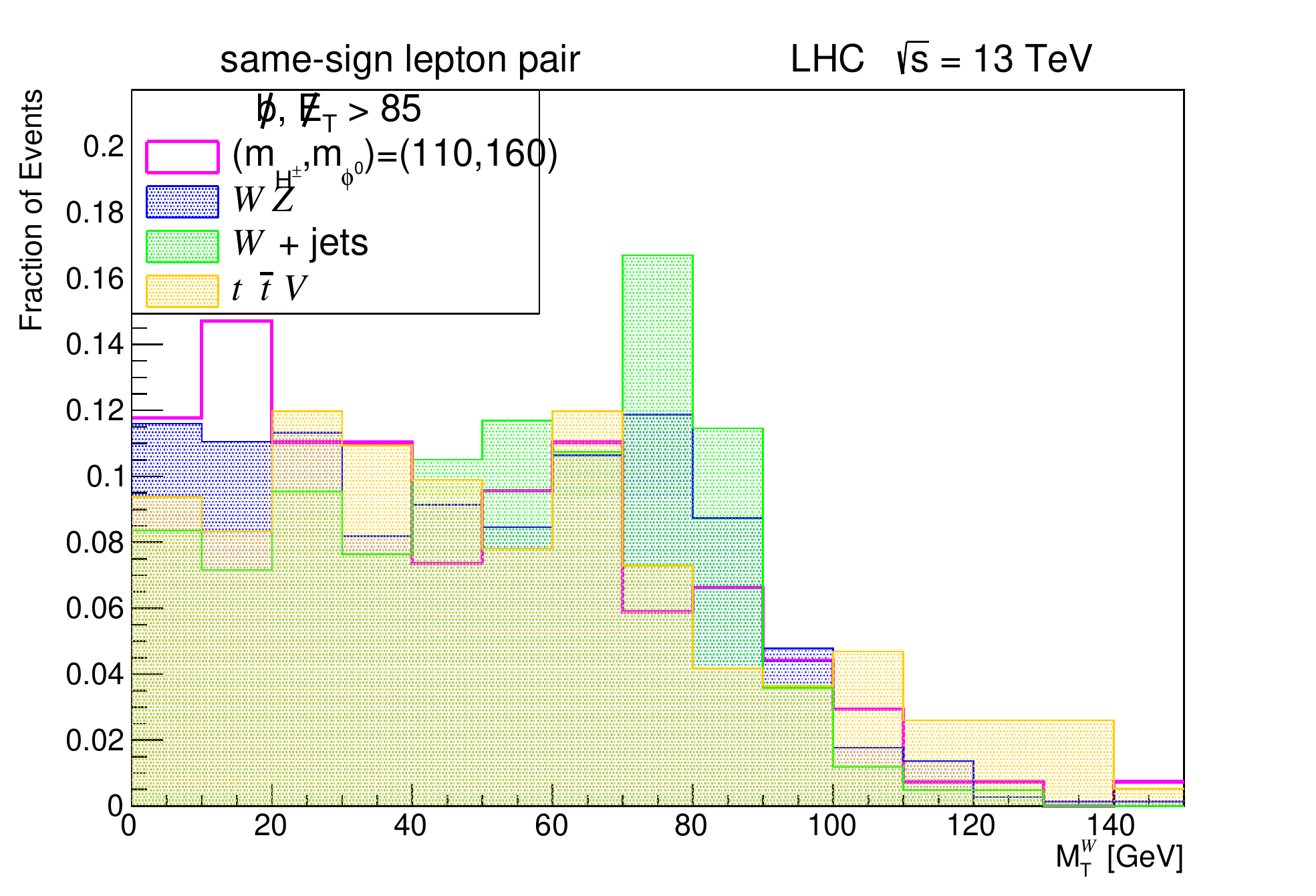}
        \caption{}
    \end{subfigure} 
    \caption{Distributions of $\Delta\phi_{\ell\ell}$ and $m_{\text{T}}^W$ for the SS lepton pair signal (Sec.\ref{2ssl}) and the main SM backgrounds after pre-selections. The signal benchmark is $(m_{H^{\pm}},m_{\phi^0})=(110,160)$ GeV with $r=1/5,\epsilon_V=0.1$ The preselections are SS $\mu\mu$ or $\mu\tau_h$, $b$-veto and $\met>85$ GeV.}
\label{ssl0b}
\end{figure*}

To combat the $WZ$ and $W+$jets backgrounds, we look at the transverse mass of the $W$:
\[
m_{\text{T}}^W
\equiv
\sqrt{2p_T^{\ell} \slashed{p}_T(1-\cos\Delta\phi_{\ell,\slashed{p}_T} )}.
\] 
Since there are two leptons, we reconstruct $m_{\text{T}}^W$ for both of them and take the smaller one to be ${m_{\text{T}}^W}$.
Based on the kinematic distributions plotted in Figure \ref{ssl0b}, the final selections comprise $7>N_j>2$, $\Delta\phi_{\ell\ell} > 2.1$ and $|{m_{\text{T}}^W}-m_W|>5$ GeV. 
Table \ref{signal160_110} gives the yields of the signal and background processes assuming an integrated luminosity of 300 fb$^{-1}$. 
To get 5$\si$, an integrated luminosity of 600fb$^{-1}$ is required.

\begin{table}
\centering
\begin{adjustbox}{max width=\textwidth}
\begin{tabular}{ l  c  c  c  c }
\hline
Signal and SM processes & $\si$(fb) & initial@300fb$^{-1}$ &pre-selection & final selection\\
\hline
 $\phi^0H^{\pm}\to(H^{\pm}W^{\mp})H^{\pm},\, H^{\pm}\to \tau\nu$ &40&$1.2\times10^4$ &116 &61 \\
\hline\hline
$W^{\pm}Z\to(\ell^{\pm}\nu)(\ell^+\ell^-)$ &1300&$3.9\times10^5$ & 599 & 116\\

$ZZ,\,Z\to \ell^+\ell^-$ &124&$3.7\times10^4$ &35 &14  \\

$t\bar{t}V$ &900&$2.7\times10^5$ &186 &49  \\

$VVV$ &440&$1.3\times10^5$ &101 & 25 \\

 $V+$jets with $V$ leptonically decay (FNP) &$3.2\times10^7$&$1.1\times10^{10}$& 644&63 \\

semi-/di-leptonic $t\bar{t}$ (FNP) &$4.0\times10^5$ &$1.2\times10^8$ & 96& 21\\
 
 \hline
SM Total & - & - & -  &288\\
\hline
\end{tabular}
\end{adjustbox}
\caption{Signal and background yields for the same-sign leptons assuming an integrated luminosity $\mathcal{L}= 300\, \text{fb}^{-1}$. To estimate the FNP leptons, we use a flat fake rate to be $\sim10^{-4}$. The signal benchmark is that $m_{H^{\pm}}=110$ GeV, $m_{\phi^0}=160$ GeV and $r=1/5, \epsilon_V=0.1$. The preselections are SS $\mu\mu$ or $\mu\tau_h$, $b$-veto and $\met>85$ GeV. The final selections are $7>N_j>2$, $\Delta\phi_{\ell\ell} > 2.1$ and $|{m_{\text{T}}^W}-m_W|>5$ GeV.}
\label{signal160_110}
\end{table}

%
\section{Conclusions} \label{sec:conclusions}
In this paper we considered the phenomenology of a 2-Higgs doublet model
where the additional Higgs bosons are almost inert.
This means that there is an approximate $\ZZ$ symmetry that ensures that
there is a Standard Model-like Higgs boson mass eigenstate whose VEV
is dominantly responsible for the masses of Standard Model vector bosons and
fermions.
This fully explains the agreement of the couplings of the observed 125~GeV
Higgs boson, while allowing the additional Higgs bosons to be light
and therefore kinematically accessible at the LHC.
The phenomenology of this kind of model is very distinctive.
The $\ZZ$ odd Higgs bosons are pair produced by electroweak interactions,
and undergo cascade decays with the heaviest Standard Model states at
the end of the decay chain.
\begin{figure}[t]
\centering
\includegraphics[scale=1]{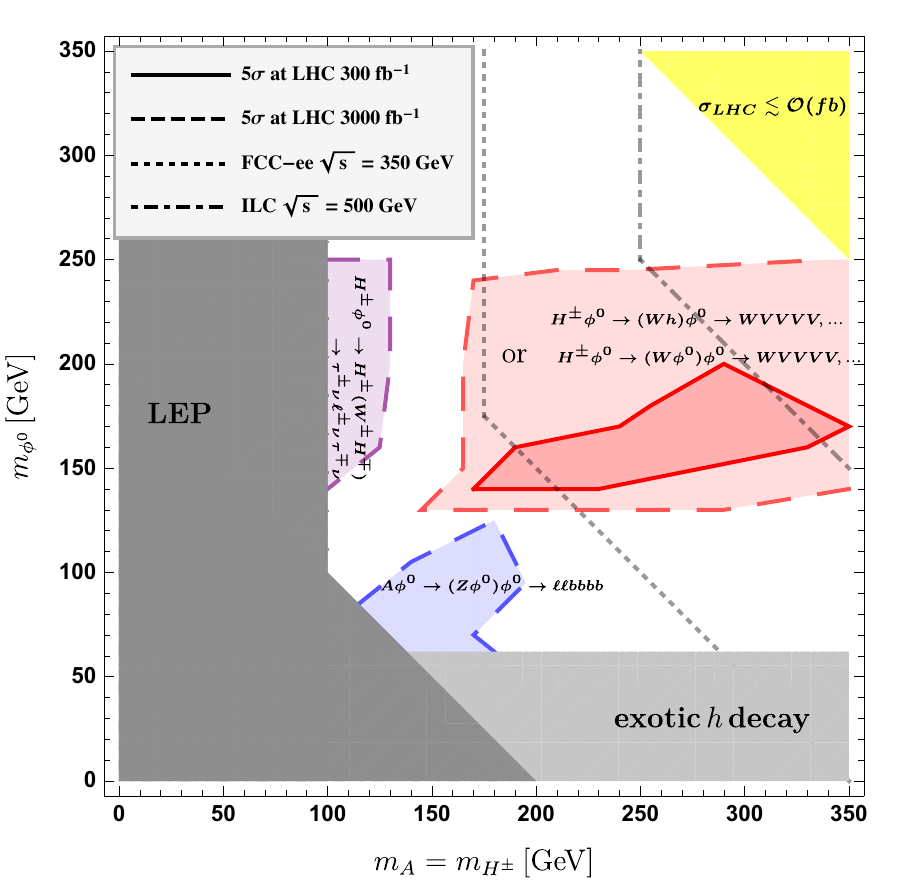}
\caption{The dashed (solid) lines are the $5\si$ reach for an integrated luminosity of 3000fb$^{-1}$ (300fb$^{-1}$) at LHC Run II. For the OSSF$\ell$-plus-3$b$ (blue) search, $r\equiv\epsilon_f/\epsilon_V=5$ is chosen. For the 2 SS$\ell$ (purple) and the 3$\ell$ off $Z$ (red) searches, $r=1/5$ is chosen.
The regions for LEP, FCC-ee, and ILC are the kinematically available regions, so they correspond to the maximal possible reach.}
\label{3000limit}
\end{figure}

\begin{table}
\centering
\begin{adjustbox}{max width=\textwidth}
\begin{tabular}{ |c|l|l|l| }
\hline
Signal  & Main Decay Modes 
& Final States &$\mathcal{L}_{5\si}$(fb$^{-1}$)\\
\hline

\multirow{2}{*}{$\phi^0 A$}

&$(b\bar{b})(\phi^0Z^*)\to(b\bar{b})(b\bar{b}\ell^+\ell^-)$ 
& OSSF+3$b$ &$300$ (Sec.~\ref{ossf_3b})\\ \cline{2-4}
&$(WW^*)(\phi^0Z^*)\to(WW^*)(WW^*Z^*)$
& 3 leptons& 300 (Sec.~\ref{3lep}) \\ \cline{2-4}
\hline

\multirow{2}{*}{$\phi^0H^{\pm}$}

&$(b\bar{b})(\phi^0W^*)\to(b\bar{b})(b\bar{b}\ell^+\nu)$ 
&1$\ell+3b$ & killed by W+jets\\ \cline{2-4}
&$(WW^*)(\phi^0W^*)\to(WW^*)(WW^*W^*)$ 
& 3 leptons &300 (Sec.~\ref{3lep}) \\ \cline{2-4}
\hline

\multirow{2}{*}{$AH^{\pm}$}

&$(\phi^0Z^*)(\phi^0W^*)\to(b\bar{b}Z^*)(b\bar{b}W^*)$ 
& 2 SSL$+3b$& killed by $t\bar{t}$  \\ \cline{2-4}
&$(\phi^0Z^*)(\phi^0W^*)\to(WW^*Z^*)(WW^*W^*)$ 
& 3 leptons &300 (Sec.~\ref{3lep})\\ \cline{2-4}
\hline

\multirow{2}{*}{$H^+H^-$}

&$(\phi^0W^*)(\phi^0W^*)\to(b\bar{b}W^*)(b\bar{b}W^*)$ 
&2$\ell+3b$ & killed by $t\bar{t}$, Z+jets\\ \cline{2-4}
&$(\phi^0W^*)(\phi^0W^*)\to(WW^*W^*)(WW^*W^*)$ 
& 3 leptons  &300 (Sec.~\ref{3lep})\\ 
\hline

\end{tabular}
\end{adjustbox}
\caption{Plausible channels assuming that $m_A=m_{H^{\pm}}>m_{\phi}$ and that $A, H^{\pm}$ undergo electroweak cascade decays. SSL means same-sign leptons. OSSF means opposite-sign same-flavor lepton pair.}
\label{triangle_low}
\end{table}

\begin{table}
\centering
\begin{adjustbox}{max width=\textwidth}
\begin{tabular}{ |c|l|l|l| }
\hline
Signal  & Main Decay Modes 
& Final States &$\mathcal{L}_{5\si}$(fb$^{-1}$)\\
\hline

\multirow{3}{*}{$\phi^0 A$}

&$(AZ^*)A\to(b\bar{b}\ell^+\ell^-)(b\bar{b})$ 
& OSSF+3$b$& signal $\si$ too small\\ \cline{2-4}
&$(H^{\pm}W^*)A\to(\tau\nu W^*)(b\bar{b})$
&$2\ell+2b$ &killed by $t\bar{t}$\\ \cline{2-4}
&$(H^{\pm}W^*)A\to(tbW^*)(b\bar{b})$ 
&$2\ell+3b$ &killed by $t\bar{t}$\\ \cline{2-4}
\hline

\multirow{3}{*}{$\phi^0H^{\pm}$}

&$(H^{\pm}W^*)H^{\pm}, \,H^{\pm}\to \tau^{\pm}\nu$ 
& 2SSL &2250 (Sec.~\ref{2ssl}) \\ \cline{2-4}
&$(H^{\pm}W^*)H^{\pm}\to(tb W^*)(t\bar{b})$  
& 2SSL+2$b$ &signal $\si$ too small\\ \cline{2-4}
&$(AZ^*)H^{\pm}\to(b\bar{b}Z^*)(\tau^{\pm}\nu)$  
& 2SSL+2$b$ &signal $\si$ too small \\ \cline{2-4}

\hline

\end{tabular}
\end{adjustbox}
\caption{Plausible channels assuming that $m_A=m_{H^{\pm}}<m_{\phi^0}$ and that $\phi^0$ undergoes electroweak cascade decays. SSL means same-sign lepton pairs. OSSF means opposite-sign same-flavor lepton pair.}
\label{triangle_up}
\end{table}

\begin{table}
\centering
\begin{adjustbox}{max width=\textwidth}
\begin{tabular}{ |c|l|l|l| }
\hline
Signal  & Main Decay Modes 
& Final States &$\mathcal{L}_{5\si}$(fb$^{-1}$) \\
\hline

\multirow{3}{*}{$A\phi^0$}

&$(b\bar{b})(b\bar{b})$ 
&$4b$ &killed by QCD\\ \cline{2-4}
&$(Z^*h)(WW^*)$\textrightarrow$(Z^*b\bar{b})(WW^*)$ 
&2SSL+2b & killed by $t\bar{t}$, $\si$ too small \\ \cline{2-4}
&$(Z^*h)(WW^*)$\textrightarrow$(Z^*VV^*)(WW^*)$  
&3 leptons &2000 (Sec.~\ref{3lep}) \\ \cline{2-4}
\hline

\multirow{3}{*}{$AH^{\pm}$}

&$(b\bar{b})(\tau\nu)$ 
&$1\ell+2b$& killed by W+jets\\ \cline{2-4}
&$(Z^*h)(W^*h)$\textrightarrow$(Z^*b\bar{b})(W^*b\bar{b})$  
& 2SSL+2-3b & signal $\si$ too small\\ \cline{2-4}
&$(Z^*h)(W^*h)$\textrightarrow$(Z^*VV^*)(W^*VV^*)$ 
&3 leptons &2000 (Sec.~\ref{3lep}) \\ 
\hline

\multirow{4}{*}{$\phi^0H^{\pm}$}

&$(b\bar{b})(\tau\nu)$  
&$1\ell+2b$& killed by $t\bar{t}$, W+jets\\ \cline{2-4}
&$(W^*W)(t^*\bar{b})$  
&2SSL+2b &killed by $t\bar{t}$\\ \cline{2-4}
&$(W^*W)(W^* h)$\textrightarrow$(W^*W)(W^* b\bar{b})$  
& 2SSL+2b&killed by $t\bar{t}$, $\si$ too small \\ \cline{2-4}
&$(W^*W)(W^* h)$\textrightarrow$(W^*W)(W^* VV^*)$ 
& 3 leptons&2000 (Sec.~\ref{3lep})  \\ 
\hline

\multirow{4}{*}{$H^+H^-$}

&$(cs)(\tau\nu)$ 
&$1\ell+2j$&killed by W+jets\\ \cline{2-4}
&$(\bar{t}^*b)(t^*\bar{b})$ 
&$2\ell+2b$ &killed by $t\bar{t}$, Z+jets \\ \cline{2-4}
&$(W^{+*}h)(W^{-*} h)$\textrightarrow$(W^{+*}b\bar{b})(W^{-*} b\bar{b})$ 
& 2SSL+2-3b & signal $\si$ too small \\ \cline{2-4}
&$(W^{+*}h)(W^{-*} h)$\textrightarrow$(W^{+*}VV^*)(W^{-*} VV^*)$  
& 3 leptons & signal $\si$ too small \\
\hline 

\end{tabular}
\end{adjustbox}
\caption{Plausible channels assuming that $A, H^{\pm}, \phi^0$ undergo non-cascade decays. SSL means same-sign leptons.}
\label{diagonal}
\end{table}

In this paper we initiated the exploration of the phenomenology of this class of models.
We focused on LHC searches, and showed that these are sensitive 
despite the low production cross sections.
The results of the investigation are summarized in Tables~\ref{triangle_low}, \ref{triangle_up}, and \ref{diagonal}.
The most effective searches are multi-lepton searches, but custom searches
involving leptons and $b$ jets are also effective.
Figure \ref{3000limit} summarizes our results. 
We show the $5\si$ reach for each search for an LHC integrated luminosity of 3000 fb$^{-1}$ 
(dashed) and 300~fb$^{-1}$ (solid).
We also compare the bounds with those from future $e^+e^-$ colliders,
which will be both clean in the background and efficient in producing the types of signals we study here.
We conclude that the high luminosity LHC can explore a significant region of the parameter
space of these well-motivated models.

\pagebreak

\appendix{Appendix A: Almost Inert Higgs in 2HDM} \label{sec:appendix}
The purpose of this note is to make contact with the conventions adopted in 2HDM literature. Here we are are going to use the mixing angles $\ep_V$ and $\ep_f$, where the notation is just a reminder that these angles are small.

The two Higgs doublet model (2HDM) extends the Standard Model (SM) Higgs sector by allowing two complex doublets. 
Without loss of generality, we choose to work with the Higgs basis, where only one of the doublets get a non-zero vacuum expectation value (VEV) after the electroweak symmetry breaking (EWSB). 
The fields can be parametrized around their VEVs as
\begin{equation}\label{higgsbasis}
\begin{split}
\mathcal H_1&=
\left (
  \begin{array}{c}
  G^{+} \\
 \frac{1}{\sqrt{2}}( v+h_1^0+iG^0)
  \end{array}
\right ),\\
\mathcal H_2&=
\left (
  \begin{array}{c}
  H^{+} \\
 \frac{1}{\sqrt{2}}(h_2^0+iA)
  \end{array}
\right ).
\end{split}
\end{equation}
The CP-even mass eigenstates are formed by linear combinations of $h_1^0$, $h_2^0$. Defining the mixing angle to be $\epsilon_V$,
\begin{equation}\label{thetav}
\begin{pmatrix} h \\ \phi^0 \end{pmatrix}
= \begin{pmatrix} \cos\epsilon_V & \sin\epsilon_V \\ -\sin\epsilon_V & \cos\epsilon_V \end{pmatrix}
\begin{pmatrix} h_1^0 \\ h_2^0 \end{pmatrix},
\end{equation}
where $h$ is the observed 125 GeV Higgs boson and $\phi^0$ the additional neutral scalar. 
The couplings of the neutral Higgs bosons to vector bosons are all related to $\epsilon_V$.
For example:
\begin{equation}
\begin{split}
h AZ & \propto \sin \epsilon_V,
\qquad
\phi^0 AZ  \propto \cos \epsilon_V,
\\
h H^{\mp}W^{\pm} & \propto \sin \epsilon_V,
\qquad
\phi^0 H^{\mp}W^{\pm}  \propto \cos \epsilon_V,
\\
h ZZ & \propto \cos\epsilon_V,
\qquad
\phi^0ZZ  \propto \sin\epsilon_V.
\end{split}
\end{equation}
In the limit that $\epsilon_V\to 0$, the $hZZ$ coupling becomes SM-like, and $\mathcal{H}_1$ behaves just as the SM doublet in terms of its gauge couplings.

The Yukawa sector of 2HDM can be written as
\begin{equation}
-\mathcal{L}_{yuk}=\,
\overline{Q}_{L_i}y_u^{ij}u_{R_j}\tilde{H}_u+\overline{Q}_{L_i}y_d^{ij}d_{R_j}H_d
+\overline{L}_{L_i}y_e^{ij}e_{R_j}H_l+h.c.,\\
\end{equation}
where $i$, $j$ are quark flavor indices and $H_u$, $H_d$, $H_l$ are linear combinations of $\mathcal{H}_1$ and $\mathcal{H}_2$. 
The mixing of $\mathcal{H}_1$ and $\mathcal{H}_2$ in $H_u$, $H_d$, $H_l$ can not be arbitrary, due to the fact that tree-level flavor changing neutral currents (FCNC) are observed to be very rare. To suppress FCNCs, what is conventionally done is to impose a $\ZZ$ symmetry to all the SM fermions and $H_u$, $H_d$, $H_l$. The $\ZZ$ basis is related to the Higgs basis in the following way:
\begin{equation}\label{higgstoZ2basis}
\left (
  \begin{array}{c}
 \mathcal H_1 \\
 \mathcal H_2
  \end{array}
\right ) = \left (
  \begin{array}{rr}
  \cos\beta & \sin\beta \\
  -\sin\beta & \cos\beta
  \end{array}
\right ) \left (
  \begin{array}{c}
  \Phi_1 \\
  \Phi_2
  \end{array}
\right ),
\end{equation}
where $\Phi_1\to -\Phi_1$, $\Phi_2\to+\Phi_2$ under a $\ZZ$ transformation, and $\tan\beta = \left< \Phi_2\right>_0/\left< \Phi_1\right>_0$.
Depending on how the fermions transform under $\ZZ$, there arise several `types' of 2HDM. 

The simplest version (type I) is to let all the SM fields even under $\ZZ$.
Therefore, in type I, only $\Phi_2 (\,=H_u=H_d=H_l$) can participate in the Yukawa interactions. Suppose the mixing angle between $\mathcal{H}_1$ and $\mathcal{H}_2$ that makes up $\Phi_2$ is
\begin{equation}
\epsilon_f\equiv\pi/2-\beta.
\end{equation}
Together with $\epsilon_V$, the couplings of the neutral Higgs bosons to the SM fermions can all be determined:
\begin{equation}\label{thetaf}
hf\bar{f} \propto \cos (\epsilon_f-\epsilon_V)/\cos\epsilon_f,
\qquad
\phi^0f\bar{f} \propto \sin(\epsilon_f-\epsilon_V)/\cos\epsilon_f,
\qquad
Af\bar{f} \propto \tan\epsilon_f.
\end{equation}
In the small $\epsilon_f$ limit (that corresponds to large $\tan\beta$) $\mathcal{H}_2$'s interactions with the SM fermions are suppressed, and $\mathcal{H}_1$ acts as the SM Higgs doublet in the Yukawa sector.
From Eq.~(\ref{higgstoZ2basis}) we can see that $\mathcal{H}_{1,2} = \Phi_{2,1}$ for $\epsilon_f \rightarrow 0$. In this limit, and only when all sources of $\ZZ$ breaking are zero (all $\epsilon s\rightarrow 0$) the approximate $\ZZ$-basis from Eq.~(\ref{eq:z2basis}) corresponds to the Higgs basis.

Following the conventions in \cite{Gunion:2002zf,Bernon:2015qea,Branco:2011iw,Gunion:1989we}, 
the mixing angle of the CP even states in the $\ZZ$ basis ($\Phi_1$,$\Phi_2$) is defined to be $\alpha$, where
\begin{equation}\label{alpha}
\left (
  \begin{array}{c}
  \phi^0_{\text{heavy}} \\
  \phi^0_{\text{light}}
  \end{array}
\right ) = \left (
  \begin{array}{rr}
  \cos\alpha & \sin\alpha \\
  -\sin\alpha& \cos\alpha
  \end{array}
\right ) \left (
  \begin{array}{c}
  \sqrt{2}\text{Re}\Phi^0_1-v_1 \\
  \sqrt{2}\text{Re}\Phi^0_2- v_2 
  \end{array}
\right ).
\end{equation}
Eq.(\ref{alpha}) together with Eqs.(\ref{higgsbasis}) and (\ref{higgstoZ2basis}) yield:
\begin{equation}\label{alpha-beta}
\left (
  \begin{array}{c}
  \phi^0_{\text{heavy}} \\
  \phi^0_{\text{light}}
  \end{array}
\right ) = \left (
  \begin{array}{rr}
  \cos(\alpha-\beta) & \sin(\alpha-\beta) \\
  -\sin(\alpha-\beta)& \cos(\alpha-\beta)
  \end{array}
\right ) \left (
  \begin{array}{c}
 h^0_1 \\
 h^0_2  \end{array}
\right ).
\end{equation}
Comparing Eq.(\ref{alpha-beta}) with (\ref{thetav}), we see that if 
$\phi^0_{\text{light}}$ is identified with the 125 GeV Higgs $h$, then $\epsilon_V\equiv\pi/2-(\beta-\alpha)$; if $\phi^0_{\text{heavy}}$ is identified with $h$, then $\epsilon_V\equiv-(\beta-\alpha)$.

To get an almost inert Higgs sector, both the gauge couplings and Yukawa couplings of the field are set to be SM-like, i.e.
\begin{eqnarray}
\epsilon_V\equiv \pi/2-(\beta-\alpha) \ \left[\text{or}-(\beta-\alpha)\right] \to 0,\quad
\epsilon_f\equiv \pi/2-\beta \to 0.
\end{eqnarray}
Therefore, we are interested in the large $\tan\beta$ limit of the type I 2HDM. There are very few experimental constraints in this limit.

Expanding the kinetic terms for $\mathcal H_2$, we obtain terms like
\begin{eqnarray}
\frac{1}{2}\sqrt{g^2+g'^2}Z_{\mu}(-\partial^{\mu}\phi^0A +\partial^{\mu}A\phi^0),\\
\frac{i}{2}\sqrt{g^2+g'^2}(c_W^2-s_W^2)Z_{\mu}(\partial^{\mu}H^{-}H^{+} -H^{-}\partial^{\mu}H^{+} ),\\
-\frac{ig}{2}W^+_{\mu}(\partial^{\mu}H^{-}\phi^0 -H^{-}\partial^{\mu}\phi^0)+h.c., \\
\frac{g}{2}W^+_{\mu}(-\partial^{\mu}H^{-}A+H^{-}\partial^{\mu}A)+h.c.
\end{eqnarray}
Therefore, the electroweak pair production of non-SM Higgs fields is not suppressed in this limit, which we will exploit in our search. 

\begin{table}[t]
\small
\begin{center}
\begin{tabular}{|c|c||c|c|c|c|c|c|}
\hline
\multicolumn{2}{|c|}{\multirow{2}{*}{$c\tau(\mbox{mm}$)}} & \multicolumn{6}{c|}{$m_{H^\pm}(\mbox{GeV})$}         \\ \cline{3-8} 
\multicolumn{2}{|c|}{}                  & $150$ & $170$ & $190$ & $210$ & $230$ & $250$ \\ \hline \hline
\multirow{6}{*}{$\epsilon_V$}        & $10^{-1}$   & $2.7\times 10^{-4}$& $4.2\times 10^{-5}$& $1.0\times 10^{-6}$ & $2.0\times 10^{-7}$ & $4.2\times 10^{-8}$ & $1.8\times 10^{-8}$ \\ \cline{2-8} 
  & $10^{-2}$   & $2.7\times 10^{-2}$& $4.2\times 10^{-3}$& $1.0\times 10^{-4}$ & $2.0\times 10^{-5}$ & $4.2\times 10^{-6}$ & $1.8\times 10^{-6}$\\ \cline{2-8} 
  & $10^{-3}$   & $2.7$              & $4.2\times 10^{-1}$& $1.0\times 10^{-2}$ & $2.0\times 10^{-3}$ & $4.2\times 10^{-4}$ & $1.8\times 10^{-4}$  \\ \cline{2-8} 
  & $10^{-4}$   & $2.7\times 10^{2}$ & $4.2\times 10^{1}$ & $1.0$               & $2.0\times 10^{-1}$ & $4.2\times 10^{-2}$ & $1.8\times 10^{-2}$ \\ \cline{2-8} 
  & $10^{-5}$   & $2.7\times 10^{4}$ & $4.2\times 10^{3}$ & $1.0\times 10^{2}$  & $2.0\times 10^{1}$  & $4.2$               & $1.8$ \\ \cline{2-8} 
  & $10^{-6}$   & $2.7\times 10^{6}$ & $4.2\times 10^{5}$ & $1.0\times 10^{4}$  & $2.0\times 10^{3}$  & $4.2\times 10^{2}$  & $1.8\times 10^{2}$\\ \hline
\end{tabular}
\caption{$c\tau$ in millimeters for different values of $\epsilon_V(=5\epsilon_f)$ and the charged Higgs mass ($m_{H^\pm}$). 
}
\label{tablectau}
\end{center}
\normalsize
\end{table}

Finally, in Table (\ref{tablectau}), we show the value of $c\tau$ for different values of $\epsilon_V(=5\epsilon_f)$ and the charged Higgs mass ($m_{H^\pm}$).

\vspace{10mm}

\section*{Acknowledgements}
This work was supported in part by the DOE under grant DE-SC-000999. N.N. was supported by FONDECYT (Chile) grant 3170906 and in part by Conicyt PIA/Basal FB0821.

\end{document}